\documentclass[11pt,a4paper]{article}
\usepackage{amssymb,amsfonts,amsmath,amsthm}
\usepackage[latin1]{inputenc}
\usepackage[english]{babel}
\usepackage[left=2.7cm,right=2.7cm,bottom=2.67cm,top=2.67cm]{geometry}
\usepackage{dsfont}
\usepackage{mathrsfs}
\usepackage{natbib}
\usepackage{enumerate}
\usepackage{color}
\RequirePackage[OT1]{fontenc}
\usepackage{graphicx}
\usepackage{multirow}
\usepackage{multicol}
\usepackage{subfigure}
\usepackage{placeins}
\usepackage{rotating}
\usepackage[flushleft]{threeparttable}
\usepackage{booktabs}
\usepackage{arydshln} 
\usepackage[normalem]{ulem}
\usepackage{url}
\usepackage{setspace}
\usepackage{pdfpages}

\graphicspath{{figures/}}

\theoremstyle{plain}

\theoremstyle{definition}

\newcommand{\E}{\mathds{E}}
\newcommand{\F}{\mathscr{F}}
\newcommand{\R}{\mathds{R}}

\newcommand{\Z}{\mathds{Z}}

\newcommand{\bs}[1]{\boldsymbol{#1}}

\newcommand{\var}{\mathrm{Var}}

\long\def\sfootnote[#1]#2{\begingroup%
\def\thefootnote{\fnsymbol{footnote}}\footnote[#1]{#2}\endgroup}

\def\bfootnote{\xdef\@thefnmark{}\@footnotetext}

\begin{document}
\pagestyle{myheadings} 
\markboth{MARMA models}{G. Pumi, D.H. Matsuoka, T.S. Prass and B.G. Palm}

\thispagestyle{empty}
{\centering
\Large{\bf A Matsuoka-Based GARMA Model for Hydrological Forecasting: Theory, Estimation, and Applications}\vspace{.8cm}\\
\normalsize{ {\bf Guilherme Pumi${}^{\mathrm{a,}}$\sfootnote[1]{Corresponding author. This Version: \today},\let\thefootnote\relax\footnote{\hskip-.3cm$\phantom{s}^\mathrm{a}$Mathematics and Statistics Institute and Programa de P\'os-Gradua\c c\~ao em Estat\'istica - Universidade Federal do Rio Grande do Sul, Brazil. \\
$\phantom{s}^\mathrm{b}$Department of Mathematics and Natural Sciences - Blekinge Institute of Technology, Sweden.
} Danilo Hiroshi Matsuoka${}^\mathrm{a}$, Taiane Schaedler Prass${}^\mathrm{a}$ and Bruna Gregory Palm${}^\mathrm{b}$
 \\
\let\thefootnote\relax\footnote{E-mails: guilherme.pumi@ufrgs.br (Pumi),  danilomatsuoka@gmail.com (Matsuoka). taiane.prass@ufrgs.br (Prass), bruna.palm@bth.se (Palm)}
\let\thefootnote\relax\footnote{ORCIDs: 0000-0002-6256-3170 (Pumi); 0000-0002-9744-8260 (Matsuoka); 0000-0003-3136-909X (Prass); 0000-0003-0423-9927 (Palm).}}\\
\vskip.3cm
}}
\begin{abstract}
Time series in natural sciences, such as hydrology and climatology, and other environmental applications, often consist of continuous observations constrained to the unit interval $(0,1)$. Traditional Gaussian-based models fail to capture these bounds, requiring more flexible approaches. This paper introduces the Matsuoka Autoregressive Moving Average (MARMA) model, extending the GARMA framework by assuming a Matsuoka-distributed random component taking values in $(0,1)$ and an ARMA-like systematic structure allowing for random time-dependent covariates. Parameter estimation is performed via partial maximum likelihood (PMLE), for which we present the asymptotic theory. It enables statistical inference, including confidence intervals and model selection. To construct prediction intervals, we propose a novel bootstrap-based method that accounts for dependence structure uncertainty.  A comprehensive Monte Carlo simulation study assesses the finite sample performance of the proposed methodologies, while an application to forecasting the useful water volume of the Guarapiranga Reservoir in Brazil showcases their practical usefulness.
\vspace{.2cm}

\vspace{.2cm}

\end{abstract}
\textbf{Statements and Declarations:} The authors declare that they have NO affiliations with or involvement in any
organization or entity with any financial interests in the subject matter or materials discussed in this manuscript.

\section{Introduction}
Most time series appearing in natural sciences, including hydrology, climatology, and other environmental applications, consist of observations that are serially dependent over time. At this point, it has been recognized that Gaussian time series models, such as the  classical autoregressive moving average models (ARMA), are too restrictive for many applications, especially in hydrology. As a consequence, the interest in non-Gaussian time series modeling has grown considerably.
This is most noticeable for time series supported on the open unitary interval $(0,1)$, such as  relative humidity, the incidence of COVID19 in some particular region (per 10,000 inhabitants, say), useful water volume of a reservoir, etc. Since models for random variables taking values on the real line may provide forecasts outside the range of plausible values for the phenomenon under study, as in \cite{grande}, specific models are needed to accommodate the natural bounds in the data.

Our motivation for this paper lies in a hydrological empirical problem, which involves modeling and forecasting the monthly useful water volume (UV) of the Guarapiranga Reservoir, located at the border between Itapecerica da Serra and Embu-Gua\c{c}u, SP, Brazil. Accurately modeling the useful water volume of a reservoir is crucial for effective water resource management.  Such modeling supports informed decision-making and is essential for successfully planning and implementing water management strategies. Forecasting plays a crucial role in this context; however, obtaining prediction confidence intervals for the forecasts is equally important. These intervals provide valuable information to quantify uncertainty, which is critical for water management.

One approach capable of handling double-bounded time series that has gained attention in the literature over the last decade is the so-called GARMA (generalized ARMA) models discussed in \cite{Benjamin2003}. In a few words, GARMA modeling merges the strengths of ARMA modeling into a generalized linear model (GLM) framework, yielding a class of very flexible models that can be easily tailored to accommodate a wide variety of structures, including non-gaussianity, bounds, asymmetries, etc. GARMA are classified as observation-driven \citep{cox1981} models and, as such, are defined by two components: the random component, which specifies the probability structure (conditional distributions) for the model, and the systematic component, which is responsible for the dependence structure present in the model.

In practice, original GARMA models only considered random components of the canonical exponential family and an ARMA-like systematic component to model the conditional mean. However, as the literature on GARMA models for continuous double-bounded time series has grown over the years, its scope has also grown. The interest nowadays lies in models for which the systematic component follows the usual approach of GLM with an additional dynamic term of the form
\begin{equation*}
g(\mu_t)={\eta}_t=\bs{X}_{t}'\bs{\beta}+\tau_t,
\end{equation*}
where $g$ is a suitable link function, $\mu_t$ is some quantity of interest {related to the conditional distribution (usually the mean or median)}, $\bs{X}_t$ denotes a vector of {(possibly random and time dependent)} covariates observed at time $t$ with associated vector of coefficient $\bs\beta$  and $\tau_t$ is a term responsible to accommodate any serial dependence in $\mu_t$. The terms $\mu_t$ and $\tau_t$ vary according to the scope of the model and the intended application. For instance, the classical ARMA form and its variants are used in \cite{Rocha2009, Maior,Prass} and \cite{Bayers, Bayerseas}, while \cite{Pumi2017} and \cite{helen} apply a long-range-dependent ARFIMA specification. More exotic non-linear specifications can also be found, as in the case of the Beta autoregressive chaotic models of \cite{BARC}.

Another important feature in GARMA models is the nature of $\mu_t$. Originally, $\mu_t$ denoted the model's conditional mean at time $t$, as in Beta-based models such as \cite{Rocha2009, Bayerseas, Pumi2017} and \cite{BARC} or as in models for positive time series \citep{Prass}. But other possibilities arose in the literature. For instance, for Kumaraswamy-based models \citep[the KARMA model of][]{Bayers}, $\mu_t$ denotes the model's median. For Unit-Weibull-based models \citep[the UWARMA of][]{uw}, it represents any quantile, while for models for which the conditional distribution is a member of a symmetric family of distributions without a finite first moment, $\mu_t$ denotes the model's point of symmetry \citep{helen}. This work introduces a new GARMA model for which the random component follows a Matsuoka's distribution in $(0,1)$, with systematic component models the process' conditional mean through an ARMA-like structure with possible exogenous covariates, called MARMA (Matsuoka autoregressive moving average) models.

We propose a partial maximum likelihood estimation (PMLE) to estimate model parameters, which allows for the inclusion of random time-dependent covariates in the model. Since the Matsuoka's distribution is a member of the canonical exponential family, the large sample theory for the PMLE follows from known results. This facilitates the construction of asymptotic confidence intervals, hypothesis testing and diagnostics. We provide a standard way to obtain out-of-sample forecasting for the proposed model. Since deriving closed form exact confidence intervals for out-of-sample forecasts may be impossible due to the complexity of the underlying stochastic processes, we propose a novel bootstrap-based method for obtaining such intervals. The proposed bootstrap scheme takes advantage of the iterative nature of the systematic components in observation-driven models, incorporating the dependence structure uncertainty into the construction of prediction intervals. Although presented in the context of MARMA models, the method generalizes straightforwardly to any GARMA model - the method is used in the context of KARMA model in the application.

The paper is organized as follows.  Section \ref{sec:model} introduces the proposed MARMA model. In Section \ref{sec:est}, results regarding parameter estimation are derived, while Section \ref{sec:lsi} is dedicated to large sample inference, including hypothesis test, goodness-of-fit, model selection, and forecasting, including the proposed bootstrap-based method for constructing prediction intervals. Finally, Section \ref{sec:MCS} presents a simulation study to explore the finite-sample performance of the model. In Section \ref{sec:Aplication} we present the main application of the paper, showcasing the use of the proposed model, highlighting the performance of the PMLE and the proposed bootstrap-based method for constructing prediction intervals, and providing a comparison with the KARMA model of \cite{Bayers}.  Section \ref{sec:Conclusion} concludes the paper.

\section{Matsuoka distribution and the MARMA model}\label{sec:model}
The Matsuoka distribution is a uniparametric distribution on $(0,1)$ absolutely continuous with respect to the Lebesgue measure with density given by
\begin{equation}\label{eq1}
f(x;p):=  2\sqrt{\frac{-p^3 \ln(x)}{\pi}} x^{p-1} I(x\in(0,1)),
\end{equation}
for $p>0$. We use the notation $X\sim M(p)$ to say that a random variable $X$ has the Matsuoka distribution with parameter $p>0$. It was first considered in \cite{consul}, which called it the log-gamma distribution and derived a few simple properties, such as moments and the distribution of some specific functions of independent variates from the distribution. The name log-gamma was not adopted in the literature, and to complicate matters,  two different kinds of distribution introduced by \cite{hogg} and  \cite{hell} also adopted the name log-gamma distribution, which makes the name dubious. A few years after \cite{consul}, \cite{Grassia} independently introduced the same distribution, derived its moments, and presented some real data applications. \cite{Grassia} did not provide a name for the distribution, which was later referred to as Grassia 1 distribution by \cite{griff}. The Matsuoka distribution is also a particular case of a very broad and general class of distributions introduced in \cite{ufpe}, called the Unit Gamma-G class. Despite all these works, little was known about the distribution until the work of \cite{mat}, where the authors study a semiparametric three-step approach to production frontier estimation. Although the paper focuses on a new method to estimate the production frontier, the distribution plays a central role in the large sample theory developed in the paper, so the authors also provide a miscellany of new results related to the distribution, which was then called the  Matsuoka distribution.

Density \eqref{eq1} presents a J-shape pattern for $p\leq 1$, it is unimodal for $p>1$ and is never symmetric about 1/2. The cumulative distribution function associated to \eqref{eq1}, is given by
\begin{align}\label{eq2}
F(x;p)=\frac{2}{\sqrt{\pi}}\Gamma\bigg(\frac32,-p\ln(x)\bigg)I(0<x<1)+I(x\geq1),
\end{align}
for $x\in\R$, where $\Gamma(k,t)=\int_t^\infty z^{k-1}e^{-z}dz$ for all $k>0$, is the upper incomplete gamma function \citep[see section 8.35 in][]{grad}. From \eqref{eq2}, it follows that the quantile function is given by $F^{-1}_p(0)=0$, $F^{-1}_p(1)=1$ and
\begin{equation*}
F^{-1}_p(q)=\exp\biggl\{-\frac1p\Gamma^{-1}\bigg(\frac32,\frac{q\sqrt{\pi}}2\bigg)\biggr\},
\end{equation*}
for $q\in(0,1)$, where $\Gamma^{-1}(k,x)$ denotes the inverse of the upper incomplete gamma function. The moments are given by $\E(X^k)=\big(\frac{p}{p+k}\big)^{\frac32}$, hence
\begin{equation*}
\E(X)=\bigg(\frac{p}{p+1}\bigg)^{\frac32}\qquad\text{and}\qquad\var(X)=\bigg(\frac{p}{p+2}\bigg)^{\frac32}-\bigg(\frac{p}{p+1}\bigg)^3.
\end{equation*}
The Matsuoka distribution is a member of the 1-parameter regular exponential family, in the form $f(x;p)=h(x)\exp\{\eta t(x)-a(\eta)\}$, with canonical parameter $\eta=p$, $a(\eta)= -\frac32\ln(\eta)$, $h(x)=-\frac{2\ln(x)}{x\sqrt{\pi}}I(0\leq x\leq 1)$, and natural complete sufficient statistics given by $T(X)=\ln(X)$. It can be shown that $-\ln(X)\sim \mathrm{Gamma}\big(\frac32,\frac1p\big)$, so that $\E(\ln(X))=-\frac3{2p}$.

Let $\{Y_t\}_{t\in\Z}$ be a stochastic process taking values in $(0,1)$ and let $\{\bs X_t\}_{t\in\Z}$ be a set of $r$-dimensional exogenous covariates to be included in the model. These can be either random or deterministic and time-dependent, or any combination of these. Let $\F_{t}$ denote the information ($\sigma$-field) available to the observer at time $t$, that is,   $\F_{t}:=\sigma\{\bs X_{t+1}, Y_{t},\bs X_{t},Y_{t-1},\cdots\}$, where, by convention, $\bs X_t$ denotes the observed values at time $t$ for deterministic covariates, and at time $t-1$ for stochastic ones. The proposed Matsuoka autoregressive moving average (MARMA) class of models is observation-driven for which the random component is implicitly defined by assigning  $Y_t|\F_{t-1}\sim M(p_t)$. Upon observing that  $\mu_t:=\E(Y_t|\F_{t-1})=\big(\frac{p_t}{1+p_t}\big)^{\frac32}$, we follow the GLM approach by setting
\begin{equation}\label{marma}
\eta_t:=g(\mu_t)= \alpha + \bs X_t^\prime \bs\beta + \sum_{i=1}^p \phi_i \big[ g(Y_{t-i})-\bs X_{t-i}^\prime \bs \beta\big] + \sum_{j=1}^q \theta_j r_{t-j},
\end{equation}
where $\eta_t$ is the linear predictor, $\alpha$ is an intercept, $\bs\beta=(\beta_1, \cdots,\beta_r)^\prime$ is the parameter vector related to the covariates, $\bs\phi=(\phi_1,\cdots,\phi_p)^\prime$ and $\bs\theta=(\theta_1,\cdots,\theta_q)^\prime$ are the AR and MA coefficients, respectively. The error term  in \eqref{marma} is defined in a recursive fashion by setting $r_t:=g(Y_t)-g(\mu_t)$. The proposed class of models, hereafter denoted MARMA$(p,q)$, is defined by setting $Y_t|\F_{t-1}\sim M(p_t)$ together with systematic component given by \eqref{marma}.

Observe that we can write $\mu_t=u(p_t)$, where $u(x):=\big(\frac{x}{1+x}\big)^{3/2}$ with inverse $u^{-1}(y)=\frac{y^{2/3}}{1-y^{2/3}}$, so that \eqref{marma} is defined in the traditional GLM fashion.  This is the approach of \cite{Benjamin2003} and  \cite{Fokianos2004}, but it is slightly different from the approach commonly used in GARMA-like models for which the distribution is not a member of the canonical exponential family, like the KARMA, UWARMA e $\beta$ARMA. In these cases, it is usually simpler to parameterize the distribution in terms of the measure of interest $\mu_t$. The end result is the same, though, and in the case of the traditional GARMA, it is a matter of preference.

In view of \eqref{marma}, it is clear that $\eta_t$ and $\mu_t$ are $\F_{t-1}$-measurable. Possible link functions to be applied in \eqref{marma} are the traditional logit, probit, loglog, and cloglog (complementary loglog), although parametric alternatives can also be considered, as, for instance, in \cite{Pumi2020}.

\section{Parameter estimation}\label{sec:est}

In this section, we propose the use of the partial maximum likelihood approach to parameter estimation in the context of MARMA models. Let $\{Y_t\}_{t\in\Z}$ be an MARMA$(p,q)$ model with associated $r$-dimensional covariates $\{\bs X_t\}_{t\in\Z}$. Let $\bs\gamma:=(\alpha,\bs\beta,\bs\phi,\bs\theta)^\prime \in \Omega$, where $\Omega\subset\R^{r+p+q+1}$ denotes the parameter space. Given a sample $\{(Y_t,\bs X_t)\}_{t=1}^n$, the partial log-likelihood function is given by
\begin{equation}\label{logL}
\ell(\bs \gamma)= \sum_{t=1}^{n} \ell_t(\bs\gamma),
\end{equation}
where
\begin{align*}
\ell_t(\bs\gamma)&:= \ln(2)+\frac32\ln(p_t)+\ln(\pi^{-1/2})+\frac12\ln\big(-\ln(Y_t)\big)+(p_t-1)\ln(Y_t)\\
&= \ln(2\pi^{-1/2})+\frac12\ln\big(-\ln(Y_t)\big)+\ln(\mu_t)-\frac32\ln(1-\mu_t^{2/3})+\bigg(\frac{\mu_t^{2/3}}{1-\mu_t^{2/3}}-1\bigg)\ln(Y_t),
\end{align*}
since $p_t=u^{-1}(\mu_t)$ and $\mu_t$ is specified by \eqref{marma}. The partial maximum likelihood estimator (PMLE) of $\bs\gamma$ is given by
\begin{equation*}
\hat{\bs\gamma}=\underset{\bs\gamma\in\Omega}{\mathrm{argmax}}(\ell(\bs\gamma)).
\end{equation*}

\subsection{The partial score vector}
In principle, the partial score vector $\frac{\partial \ell(\bs\gamma)}{\partial \bs\gamma}$ can be used to obtain $\hat{\bs\gamma}$ by solving the system $\frac{\partial \ell(\bs\gamma)}{\partial \bs\gamma}=\bs 0$. To obtain the partial score vector, in view of \eqref{logL}, we only need to derive $\frac{\partial \ell_t(\bs\gamma)}{\partial \bs\gamma}$. In what follows, all equalities are to be understood to hold almost surely. Observe that
\begin{align*}
\frac{\partial \ell_t(\bs\gamma)}{\partial \gamma_j}
&=\frac{\partial \ell_t(\bs\gamma)}{\partial \mu_t}\frac{\partial \mu_t}{\partial \eta_t}\frac{\partial \eta_t}{\partial \gamma_j}
=\frac1{g'(\mu_t)}\bigg(\frac{2\ln(Y_t)}{3\big(1-\mu_t^{2/3}\big)^2\mu_t^{1/3}}+ \frac1{\big(1-\mu_t^{2/3}\big)\mu_t}\bigg) \frac{\partial \eta_t}{\partial \gamma_j},
\end{align*}
where the last equality follows since $\frac{\partial\mu_t}{\partial \eta_t}=\frac1{g'(\mu_t)}$. Since $\eta_t$ given in  \eqref{marma} is exactly the same specification for the KARMA model, the derivatives $\frac{\partial \eta_t}{\partial \gamma_j}$ follow the same recursions as the ones presented in section 3 in \cite{Bayers}, namely
\begin{align}\label{equas}
\frac{\partial \eta_t}{\partial \alpha}&=1 - \sum_{j=1}^q \theta_j \frac{\partial \eta_{t-j}}{\partial \alpha},\qquad\qquad
\frac{\partial \eta_t}{\partial \beta_l}= X_{tl}-\sum_{i=1}^{p} \phi_iX_{(t-i)l}- \sum_{j=1}^q \theta_j \frac{\partial \eta_{t-j}}{\partial \beta_l},\nonumber\\
\frac{\partial \eta_t}{\partial \phi_k}&= g({Y}_{t-1})-\bs X_{t-1}'\bs\beta - \sum_{j=1}^q \theta_j \frac{\partial \eta_{t-j}}{\partial \phi_k},\qquad\text{and}\qquad
\frac{\partial \eta_t}{\partial \theta_s}=r_{t-j} - \sum_{i=1}^q \theta_i \frac{\partial \eta_{t-i}}{\partial \theta_s},
\end{align}
for $l\in\{1,\cdots,r\}$, $k\in\{1,\cdots,p\}$ and $s\in\{1,\cdots,q\}$, where $X_{tl}$ denotes the $l$-th component of $\bs X_t$. Let  $D_{\bs\gamma}$ be the $n\times(p+q+r+1)$  matrix whose $(i,j)$th elements is given by
\begin{equation*}
[D_{\bs\gamma}]_{i,j} = \frac{\partial \eta_{i}}{\partial \gamma_j},\qquad\bs h:=\bigg(\frac{\partial \ell_1(\bs\gamma)}{\partial \mu_1},\cdots, \frac{\partial \ell_n(\bs\gamma)}{\partial \mu_n}\bigg)^\prime,
\end{equation*}
and $T$ be a diagonal matrix given by
\begin{align*}
T:=\mathrm{diag}\biggl\{\frac{\partial \mu_1}{\partial \eta_{t}},\cdots,\frac{\partial \mu_n}{\partial \eta_{n}}\biggr\} = \mathrm{diag}\biggl\{\frac{1}{g'(\mu_1)},\cdots,\frac{1}{g'(\mu_n)}\biggr\}.
\end{align*}
With these definitions, the partial score vector can be compactly written as
\begin{equation*}
U(\bs\gamma) :=\frac{\partial \ell(\bs\gamma)}{\partial \bs\gamma}= D_{\bs\gamma}' T\bs h.
\end{equation*}
Observe that from \eqref{equas}, it is clear that the PMLE cannot be obtained analytically, so we have to resort to numerical optimization to accomplish that.

\subsection{Conditional information matrix}

In this section, we obtain the conditional information matrix in closed form, which will be useful later on when deriving the asymptotic properties of the partial maximum likelihood estimator for the proposed model. Equalities in this section are to be understood to hold almost surely. Let
\begin{equation*}
H_t(\bs \gamma) := -\frac{\partial^2\ell_t(\bs\gamma)}{\partial \bs \gamma \partial \bs \gamma'},\quad\text{and}\quad
H(\bs \gamma) := -\frac{\partial^2\ell(\bs\gamma)}{\partial \bs \gamma \partial \bs \gamma'}  =  -\sum_{t=1}^{n}\frac{\partial^2\ell_t(\bs\gamma)}{\partial \bs \gamma \partial \bs \gamma'} = \sum_{t=1}^nH_t(\bs \gamma).
\end{equation*}
Notice that $H(\bs \gamma)$ and $\ell(\bs\gamma)$ both depend on $n$. However, for simplicity and since no confusion will arise, we shall drop the dependence on $n$ on the notation. Let $I_n(\bs \gamma) := \E(H(\bs \gamma))$ be the information matrix corresponding to the sample of size $n$ and let
\begin{equation*}
   I^{(n)}(\bs \gamma) := -\frac{1}{n}\sum_{t = 1}^n \E \bigg(\frac{\partial^2\ell_t(\bs \gamma)}{\partial \bs\gamma \partial \bs \gamma'} \bigg)= -\frac{1}{n}\E \bigg(\frac{\partial^2\ell(\bs \gamma)}{\partial \bs \gamma \partial \bs \gamma'} \bigg),
\end{equation*}
so that $I_n(\bs \gamma ) = nI^{(n)}(\bs \gamma)$. Now, observe that
\begin{equation*}
 I^{(n)}(\bs \gamma) = -\frac{1}{n}\sum_{t = 1}^n \E \bigg( \E\bigg(\frac{\partial^2\ell_t(\bs \gamma)}{\partial \bs\gamma \partial \bs \gamma'} \bigg| \F_{t-1}\bigg)\bigg)  = \frac{1}{n}\E\big(K_n(\bs \gamma)\big),
\end{equation*}
with
\begin{equation*}
 K_n(\bs \gamma) : = -\sum_{t = 1}^n \E \bigg(\frac{\partial^2\ell_t(\bs \gamma)}{\partial \bs\gamma \partial \bs\gamma'} \Big| \F_{t-1}\bigg).
\end{equation*}
The matrix $K_n(\bs \gamma)$ is known as the conditional information matrix corresponding to the sample of size $n$.
Under some regularity conditions (see the discussion in the next section),
\begin{equation}\label{ah}
 \frac{1}{n}H(\bs \gamma) - I^{(n)}(\bs \gamma) \overset{P}{\longrightarrow} 0 \quad \text{and} \quad    \frac{1}{n}K_n(\bs \gamma) - I^{(n)}(\bs \gamma) \overset{P}{\longrightarrow} 0, \quad \text{as} \quad n\to \infty.
\end{equation}
Furthermore, $I^{(n)}(\bs \gamma) \, {\longrightarrow} \,  I(\bs \gamma)$, where
\begin{equation}\label{I}
I(\bs \gamma) = \lim_{n\to \infty} I^{(n)}(\bs \gamma) = \lim_{n\to \infty} -\frac{1}{n}\E \bigg(\frac{\partial^2\ell(\bs \gamma)}{\partial \bs \gamma \partial \bs \gamma'} \bigg),
\end{equation}
which is the analogous of the $I_1(\bs \gamma)$ matrix for i.i.d. samples. We shall derive $K_n(\bs\gamma)$ in closed form. First notice that
\begin{align*}
\frac{\partial^2\ell(\bs\gamma)}{\partial \gamma_i \partial \gamma_j} &= \sum_{t=1}^{n}\frac{\partial}{\partial \mu_t}
\left( \frac{\partial \ell_t(\bs\gamma)}{\partial \mu_t}\frac{\partial \mu_t}{\partial \eta_t} \frac{\partial \eta_t}{\partial \gamma_j}\right)
\frac{d \mu_t}{d \eta_t} \frac{\partial \eta_t}{\partial \gamma_i} \\
&= \sum_{t=1}^{n} \left[ \frac{\partial^2 \ell_t(\bs\gamma)}{\partial \mu_t^2}\frac{\partial \mu_t}{\partial \eta_t} \frac{\partial \eta_t}{\partial \gamma_j}
+ \frac{\partial \ell_t(\bs\gamma)}{\partial \mu_t}\frac{\partial}{\partial \mu_t}\left(\frac{\partial \mu_t}{\partial \eta_t} \frac{\partial \eta_t}{\partial \gamma_j} \right) \right]
\frac{d \mu_t}{d \eta_t} \frac{\partial \eta_t}{\partial \gamma_i}\,.
\end{align*}
Since $\frac{\partial \ell_t(\mu_t,\varphi)}{\partial \mu_t} =\frac3{2p_t}+\ln(Y_t)$, it follows that $\E\Big(\frac{\partial \ell_t(\mu_t,\varphi)}{\partial \mu_t} \big| \F _{t-1}\Big)=0$, and by the $\F_{t-1}$-measurability of $\mu_t$ and $\eta_t$, it follows that
\begin{equation*}
\E\bigg(\frac{\partial^2\ell_t(\bs\gamma)}{\partial \gamma_i \partial \gamma_j}\Big|\F_{t-1}\bigg)=  \E\bigg(\frac{\partial^2 \ell_t(\bs\gamma)}{\partial \mu_t^2}\Big|\F_{t-1}\bigg)\bigg[\frac{\partial \mu_t}{\partial \eta_t}\bigg]^2 \frac{\partial \eta_t}{\partial \gamma_i}\frac{\partial \eta_t}{\partial \gamma_j},
\end{equation*}
where $\frac{\partial \eta_t}{\partial \gamma_k}$ is given in \eqref{equas}. Hence, we only need to obtain $\E\Big(\frac{\partial^2 \ell_t(\bs\gamma)}{\partial \mu_t^2}\big|\F_{t-1}\Big)$.
\begin{align*}
\frac{\partial^2 \ell_t(\bs\gamma)}{\partial \mu_t^2}&=\frac{\partial}{\partial \mu_t}\bigg[\frac{2\ln(Y_t)}{3\big(1-\mu_t^{2/3}\big)^2\mu_t^{1/3}}+ \frac1{\big(1-\mu_t^{2/3}\big)\mu_t}\bigg]\\
&=\frac23\ln(Y_t)\frac{\partial}{\partial \mu_t}\Big[\big(1-\mu_t^{2/3}\big)^{-2}\mu_t^{-1/3}\Big] + \frac{\partial}{\partial \mu_t}\bigg[\frac1{\big(1-\mu_t^{2/3}\big)\mu_t}\bigg]\\
&=\frac23\ln(Y_t)\bigg[\frac{1-5\mu_t^{2/3}}{3\big(1-\mu_t^{2/3}\big)^3\mu_t^{4/3}}\bigg]
+ \frac{5\mu_t^{2/3}-3}{3\big(1-\mu_t^{2/3}\big)^2\mu_t^2}.
\end{align*}
Now, since $\E\big(\ln(Y_t)|\F_{t-1}\big)=-\frac3{2p_t}=-\frac{3(1-\mu_t^{2/3})}{2\mu_t^{2/3}}$ and $\mu_t$ is $\F_{t-1}$-measurable,
\begin{align*}
\E\bigg(\frac{\partial^2 \ell_t(\bs\gamma)}{\partial \mu_t^2}\Big|\F_{t-1}\bigg)&=-\frac{1-\mu_t^{2/3}}{\mu_t^{2/3}}\bigg[\frac{1-5\mu_t^{2/3}}{3\big(1-\mu_t^{2/3}\big)^3\mu_t^{4/3}}\bigg] + \frac{5\mu_t^{2/3}-3}{3\big(1-\mu_t^{2/3}\big)^2\mu_t^2}
=\frac{10\mu_t^{2/3}-4}{3\big(1-\mu_t^{2/3}\big)^2\mu_t^2}.
\end{align*}
By letting $E_{\mu}$ be the $n\times n$ diagonal matrix for which the $k$th diagonal elements is given by
\begin{equation*}
[E_\mu]_{k,k}:=-\E\bigg(\frac{\partial^2 \ell_k(\bs\gamma)}{\partial \mu_k^2}\Big|\F_{k-1}\bigg)=\frac{4-10\mu_k^{2/3}}{3\big(1-\mu_k^{2/3}\big)^2\mu_k^2},
\end{equation*}
and $D_{\bs\gamma}$ and $T$ as before, we obtain
\begin{equation*}
K_n(\bs\gamma)=D_{\bs\gamma}'TE_{\mu}TD_{\bs\gamma}.
\end{equation*}

\section{Large sample inference} \label{sec:lsi}
The asymptotic theory for the proposed PMLE in the context of MARMA$(p,q)$ models falls into the general theory of \cite{Fokianos2004}, since the Matsuoka distribution is a member of the exponential family in canonical form. To simplify the presentation,  consider model \eqref{marma} without any covariates and let $Y_1,\cdots,Y_n$ be a sample from an MARMA$(p,q)$ model and let $\bs Z_{t-1}:=\big(1,g(Y_{t-1}),\cdots, g(Y_{t-p}), r_{t-1},\cdots,r_{t-q}\big)'$, so that \eqref{marma} becomes $\eta_t=\bs Z_{t-1}'\bs\gamma$, where $\bs\gamma= (\alpha,\bs\phi',\bs\theta')'$. The conditions for the consistency and asymptotic normality of the PMLE for $\bs \gamma$ are presented in \cite{Fokianos2004}, which are presented here for completeness and to fix the notation:
\begin{enumerate}
\item The true parameter $\bs\gamma_0$ belongs to an open set $\Omega\subseteq\R^{p+q+1}$ and $\bs Z_{t-1}$ almost surely lies on a compact subset $\Gamma\subset \R^{p+q+1}$, such that $P\big(\sum_{t=1}^n\bs Z_{t-1}\bs Z_{t-1}'>0\big)=1$.
\item The link function $g$ is twice continuously differentiable with inverse $g^{-1}$ satisfying $\partial g^{-1}(x)/\partial x\neq 0$ and so that $\bs Z_{t-1}'\bs\gamma$ belongs almost surely in the domain of $g^{-1}$, for all $\bs\gamma\in\Omega$ and all $t$.
    \item  There is a probability measure $\lambda$ on $\R^{p+q+1}$ such that $\displaystyle{\int_{\R^{p+q+1}}} \bs v\bs v' \lambda(d\bs v)$ is positive definite and such that
\begin{equation*}
\frac1n\sum_{t=1}^n I(\bs Z_{t-1}\in A)\overset{P}{\longrightarrow} \lambda(A),\qquad \text{as $n\rightarrow\infty$, at $\bs \gamma_0$}.
\end{equation*}
\end{enumerate}
A detailed discussion on these assumptions and their implications can be found on section 5 of \cite{Fokianos2004}. In particular, condition 3 calls for a type of ergodic theorem in the sense that if $h$ is a continuous bounded function defined on $\Gamma$,
\begin{equation*}
\frac1n\sum_{t=1}^n h(\bs Z_{t-1})\overset{P}{\longrightarrow} \displaystyle{\int_{\R^{p+q+1}}} h(\bs v) \lambda(d\bs v),
\end{equation*}
which implies \eqref{ah}, with $I(\bs\gamma_0)$ positive definite and, thus, invertible. Under Assumptions 1 to 3, for large $n$ an almost surely unique PMLE $\hat{\bs\gamma}$ exists and satisfy $\hat{\bs\gamma}\overset{P}{\longrightarrow}\bs\gamma_0$. Furthermore, \eqref{ah} is satisfied and
\begin{equation}\label{clt}
\sqrt{n}(\hat{\bs\gamma}-\bs\gamma_0)\longrightarrow N_{p+q+1}\big(\bs 0, I^{-1}(\bs\gamma_0)\big),
\end{equation}
in distribution, as $n\rightarrow\infty$, where $I$ is given in \eqref{I} and $N_{m}\big(\bs 0, \Sigma \big)$ denotes the $m$-variate normal distribution with mean vector $\bs 0=(0,\cdots,0)'\in\R^{m}$ and variance-covariance matrix $\Sigma$. The proof of these results is based on a careful analysis of the asymptotic behavior of the conditional information matrix along with a central limit theorem for the properly normalized partial score vector $U(\bs\gamma)$. Further details can be found in \cite{Fokianos1998, Fokianos2004}.

\subsection{Hypothesis testing}
\label{s:hp}
Let $Y_1,\cdots,Y_n$ be a sample from an MARMA$(p,q)$ model with covariates $\bs X_1,\cdots,\bs X_n$, with $\bs X_k\in\R^r$,  and let $\bs\gamma_0:=(\gamma_1^0,\cdots,\gamma_{p+q+r+1}^0)'$ and $\hat{\bs\gamma}:=(\hat\gamma_1,\cdots,\hat\gamma_{p+q+r+1})'$ denote the true and the obtained PMLE estimate. The central limit theorem \eqref{clt} provides a familiar framework for the construction of asymptotic confidence intervals and test statistics similar to those used in the i.i.d. context, since the approximation $\sqrt{K_n(\hat{\bs\gamma})^{jj}}(\hat\gamma_j-\gamma_j^0)\approx N(0,1)$ holds for all large enough $n$, where $K_n(\hat{\bs\gamma})^{jj}$ denotes the $j$th diagonal element of $K_n(\hat{\bs\gamma})^{-1}$. Level $\delta$ confidence intervals for $\gamma_j^0$ can be obtained straightforwardly as $\hat\gamma_j\pm z_{1-\delta/2}/\sqrt{K_n(\hat{\bs\gamma})^{jj}}$, where $z_{1-\delta/2}$ is the $(1-\delta/2)$th quantile from a standard normal distribution. Tests of the form $H_0:\gamma_j=\gamma_j^\ast$ for some prespecified $\gamma_j^\ast$ can be carried on using the following Wald's $z$ statistics
\begin{equation*}
z=\frac{\hat\gamma_j-\gamma_j^\ast}{\sqrt{K_n(\hat{\bs\gamma})^{jj}}},
\end{equation*}
which  is approximately standard normal distributed under the null hypothesis and for all sufficiently large $n$. Other traditional tests such as Rao's score, likelihood ratio, among others, are constructed analogously and asymptotically follow the same distribution as their counterparts under independence. See \cite{Fahrmeir1987} and section 6 of \cite{Fokianos2004}.

\subsection{Residuals and goodness-of-fit}\label{gof}
Residual analysis and goodness-of-fit tests are crucial for any time series analysis. However, contrarily to ARMA models, the error term in specification \eqref{marma} is constructed iteratively so that no information about its distribution is available. In fact, since $r_t=g(Y_t)-g(\mu_t)$, unless $g$ is the identity function, $\E(r_t)$ cannot be computed and it is likely to be non-zero. Hence, the obvious candidate for a residual analysis $\hat r_t=g(Y_t)-g(\hat\mu_t)$, where $\hat\mu_t$ is obtained from the PMLE estimate $\hat{\bs\gamma}$, is not a good one. Observe, however, that since $\mu_t=\E(Y_t|\F_{t-1})$, the simple residuals defined by $e_t:=Y_t-\mu_t$ are such that $\{(e_t,\F_{t-1})\}_{t\in\Z}$ is a martingale difference sequence. Hence, if the MARMA model is well specified, $\hat e_t:=Y_t-\hat\mu_t$ should behave as a martingale difference with respect to $\F_{t-1}$. This can be tested using a martingale difference test. Another commonly used approach is based on the so-called quantile residuals, defined as
\begin{equation}\label{qres}
e_t^{(q)}:=\Phi^{-1}(F(Y_t|\F_{t-1})),
\end{equation}
where $F(\cdot|\F_{t-1})$ denotes the cumulative distribution function associated with the model's random component and $\Phi^{-1}$ denotes the standard normal quantile function. In the present setting, when the model is correctly specified, the quantile residuals obtained from \eqref{qres} after plugin-in the PMLE on \eqref{eq2}, should follow a standard normal distribution \citep[see lemma 2 in][]{kall}. These goodness-of-fit procedures will be further explored in the simulations.

\subsection{Model selection and forecasting} \label{msf}

Forecasting follows the same approach as other GARMA-like models such as the KARMA, UWARMA and $\beta$ARMA models. Let $Y_1,\cdots, Y_n$ be a sample of an MARMA$(p,q)$ model with associated covariates $\bs X_1,\cdots,\bs X_n$. To obtain $h$-step ahead forecasts  $\hat Y_{n+1}, \cdots, \hat Y_{n+h}$, we assume that future values $\bs X_{n+1},\cdots,\bs X_{n+h}$ are available or can be obtained (by forecasting, for instance). With the PMLE $\hat{\bs\gamma}$ in hand, starting at $t=1$, we recursively obtain
\begin{equation}\label{rec}
\hat \eta_{t}  = \hat\alpha + \hat{\bs X}_t'\hat{\bs\beta} + \sum_{i=1}^p  \hat\phi_i\bigl[g(\hat Y_{t-i})-\hat{\bs X}_{t-i}'\hat{\bs\beta}\bigr] + \sum_{k=1}^q \hat \theta_k \hat r_{t-k},
\end{equation}
with $\hat \mu_t  = g^{-1}(\hat \eta_t)$, for $t \geq 1$, $\hat r_t = (g(\hat Y_t) - \hat \eta_t)I(1 \leq t \leq n)$,
\begin{equation}\label{cases}
    \hat Y_t = \begin{cases}
      g^{-1}(0), & p > 0, \ t < 1,\\
   Y_t, & 1 \leq  t \leq n,\\
  \hat \mu_t, & t > n,
   \end{cases}\qquad\text{and}\qquad
   \hat{\bs X}_t = \begin{cases}
  \frac{1}{p}\sum\limits_{i = 1}^p \bs X_i, & p >0, \ t < 1,\\
\bs X_t, & t \geq 1.
\end{cases}
\end{equation}
The sequence $\hat\mu_{1},\cdots,\hat\mu_{n}$ are the in-sample forecasted values whereas the $h$-step ahead forecasted values are obtained by setting $\hat Y_{n+k}=\hat\mu_{n+k}$ for $k\in\{1,\cdots,h\}$.

Different values for $\hat Y_t$ and $\hat{\bs X}_t$ when $t<0$ in \eqref{cases} may be chosen. Our experiments suggest that as long as these initial values are ``reasonable'' and the sample size is not too small, the effects of these default values on the forecast are negligible. Values of $r_t$ for $t\notin\{1,\cdots,n\}$ were taken as 0, but any reasonable values can be used with negligible impact on the forecasts.

In view of \eqref{ah} and \eqref{clt}, the delta method allows for the construction of approximate level $\delta$ in-sample forecasting interval through \citep{Fokianos2004}
\begin{equation*}
CI(\mu_t;\delta) = \hat\mu_t \pm z_{1-\frac\delta2}\sqrt{\frac{\bs Z_{t-1}'K_n(\hat{\bs\gamma})^{-1}\bs Z_{t-1}}{ng'(\hat\mu_t)^2}}.
\end{equation*}
Out-of-sample forecasting intervals in the context of $\beta$ARMA models was studied in \cite{bruna}, where 5 different approaches, 3 of them bootstrap-based, are investigated. Although the methods presented there can be in principle adapted to the present context, here we shall consider a new approach based on the recurrence nature of \eqref{rec}. The proposed approach is as follows. Let $Y_1,\cdots,Y_n$ be a sample from an MARMA$(p,q)$ model and $\hat{\bs\gamma}$ be the PMLE. Suppose we are interested in obtaining $h$-step ahead forecasting intervals for $Y_t$. The idea is to generate $m$ bootstrap samples for $Y_{n+1}, \cdots,Y_{n+h}$ from an MARMA$(p,q)$ model considering $\hat{\bs\gamma}$ as parameter. To accomplish that, we start by reconstructing the sequences $\hat{\mu}_1,\cdots,\hat{\mu}_{n+1}$ and $\hat r_{1}, \cdots, \hat r_n$ using the PMLE $\hat{\bs\gamma}$ through \eqref{rec} as before. For each $b\in\{1,\cdots,m\}$, the algorithm initiate by sampling $\hat Y_{n+1}^{(b)}$ from a Matsuoka distribution with parameter $\hat\mu_{n+1}$, which can be obtained through \eqref{rec} and update $\hat r_{n+1}^{(b)} = g(\hat Y^{(b)}_{n+1})-g(\hat \mu_{n+1})$. Then we perform the following two steps sequentially, for $k \in\{2,\cdots,h\}$:
\begin{enumerate}
\item Update $\hat\mu_{n+k}^{(b)}$ through \eqref{rec} using $Y_1,\cdots, Y_n$, $\hat Y_{n+1}^{(b)}, \cdots,\hat Y_{n+k-1}^{(b)}$ and $\hat r_1,\cdots, \hat r_n,$ \linebreak $\hat r_{n+1}^{(b)}, \cdots,\hat r_{n+k-1}^{(b)}$.
\item  Sample $\hat Y^{(b)}_{n+k}$ from a Matsuoka distribution with parameter $\hat\mu_{n+k}^{(b)}$ and update \linebreak $\hat r_{n+k}^{(b)} = g(\hat Y^{(b)}_{n+k})-g(\hat \mu^{(b)}_{n+k})$.
\end{enumerate}
From these steps, we obtain a collection $\big\{\hat Y^{(b)}_{n+1},\cdots, \hat Y^{(b)}_{n+h}\big\}_{b=1}^m$ of bootstrap samples. For each $k\in\{1,\cdots,h\}$, a level $\delta$ confidence interval for $Y_{n+k}$ is obtained from the $(1-\delta/2)$th and $\delta/2$th sample quantiles calculated from $\hat Y^{(1)}_{n+k}, \cdots,\hat Y^{(m)}_{n+k}$. We explore the finite-sample properties of the proposed approach in the simulations presented in Section \ref{sec:MCS}.

Model selection in the context of GARMA model may be conducted using information criteria such as the AIC, BIC and HQC, which are calculated as per usual based on the maximized partial likelihood. Bayesian approaches via Reversible Jump Markov Chains have also been considered in the literature \citep{casarin}, but we shall not delve into this matter in this paper.

\section{Monte Carlo Simulation}\label{sec:MCS}

In this section we explore the finite-sample performance of the proposed PMLE approach for parameter estimation in MARMA models. Our goal is to study point and interval estimation and residual analysis. The latter will be explored using the approaches delineated in Section \ref{gof}. The simulation was carried out using R version 4.3.1 \citep{R}.
\subsection{Point estimation}
In this section, we conduct a simulation study to assess the finite-sample performance of the PMLE for MARMA models, focusing on point estimation. We evaluate the accuracy and reliability of the estimator across different parameter configurations and sample sizes, considering both mean and median estimates along with their dispersion. Particular attention is given to potential biases and skewness in small samples and the impact of parameter values on estimation performance. Additionally, we investigate the asymptotic normality of the PMLE, developed in Section \ref{sec:lsi}, by analyzing the distributional behavior of the estimates as the sample size increases. The results provide insights into the strengths and limitations of PMLE in practical settings.
\subsubsection*{Data generating process}
\begin{figure}[!ht]
\centering
\includegraphics[width=0.7\textwidth]{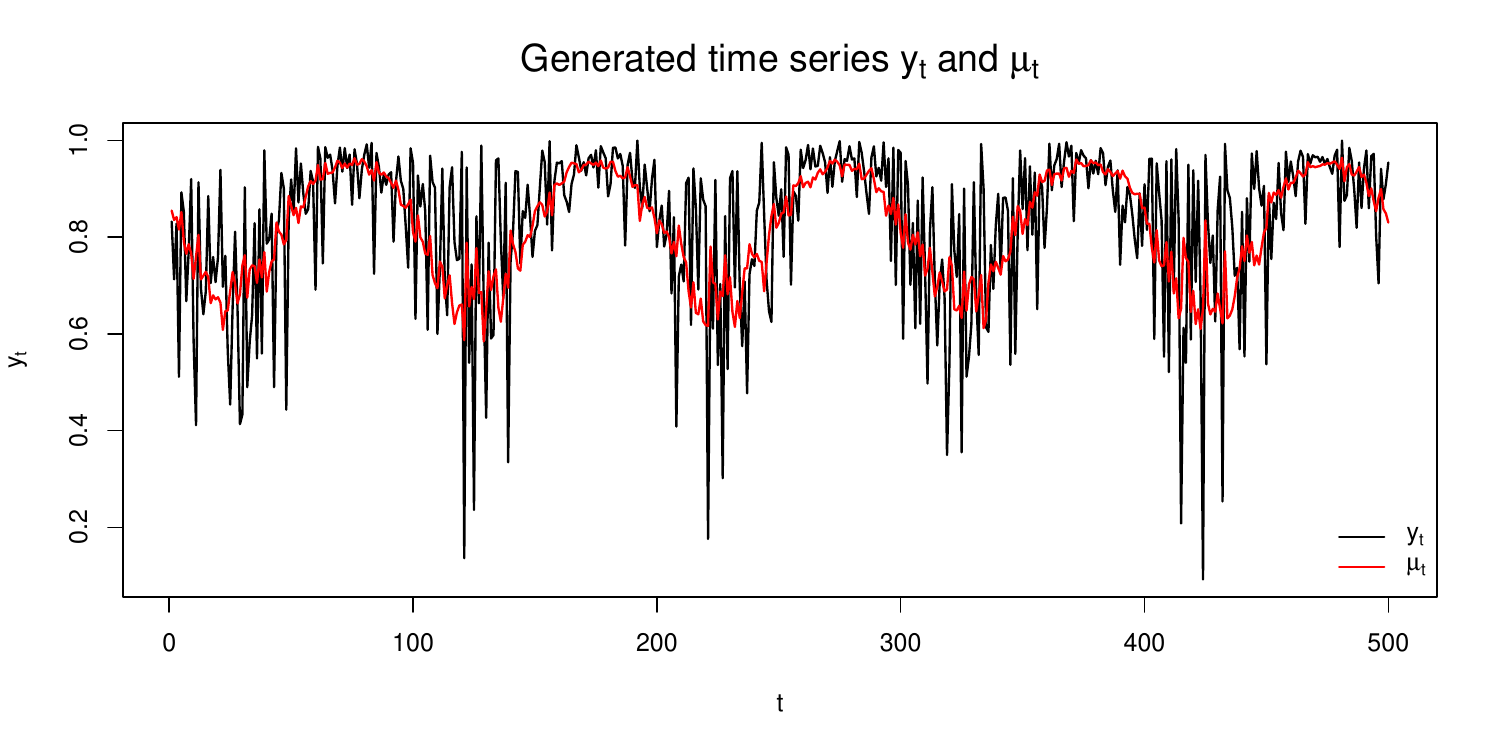}
\caption{A typical example of a time series considered in the simulation study.
The plot was generated considering $n=500$, $\alpha = 0.5$, $\beta=-0.5$, $\phi=0.2$, $\theta=-0.4$.} \label{ref}
\end{figure}

We generate samples of size $n\in\{100,200,500\}$ from an MARMA$(1,1)$ model with parameters $\alpha\in\{0.5,1\}$, $(\phi,\theta) \in\big\{(0.2,-0.8),(-0.8,0.2),(-0.4,-0.2),(0.4,0.2)\big\}$, considering a sinusoidal covariate given by $X_t=\sin\big(\pi t/50\big)$ with coefficient $\beta=-0.5$.  As link function, we consider $g(x) = \log(-\log(1-x))$, known in the literature as cloglog. Under these specifications, the underlying model is given by
\begin{equation*}
\eta_t := g(\mu_t)= \alpha  -0.5 \sin\big(\pi t/50\big) + \phi \big[ g(Y_{t-1}) + 0.5 \sin\big(\pi (t-1)/50\big) \big] + \theta r_{t-1},
\end{equation*}
 A size 100 burn-in is applied to generate the time series. A total of 1,000 replicas of each scenario were generated. Routines to sample from a MARMA$(p,q)$ process and to perform estimation via PMLE will be available in the forthcoming version of R package \texttt{BTSR} \citep{btsr} to be released soon.  For reference, a typical realization of such a process is presented in Figure \ref{ref} along with the respective $\mu_t$.
\FloatBarrier
\subsubsection*{Simulation results: point estimation}
\begin{figure}[!ht]
\centering
\includegraphics[width=0.9\textwidth]{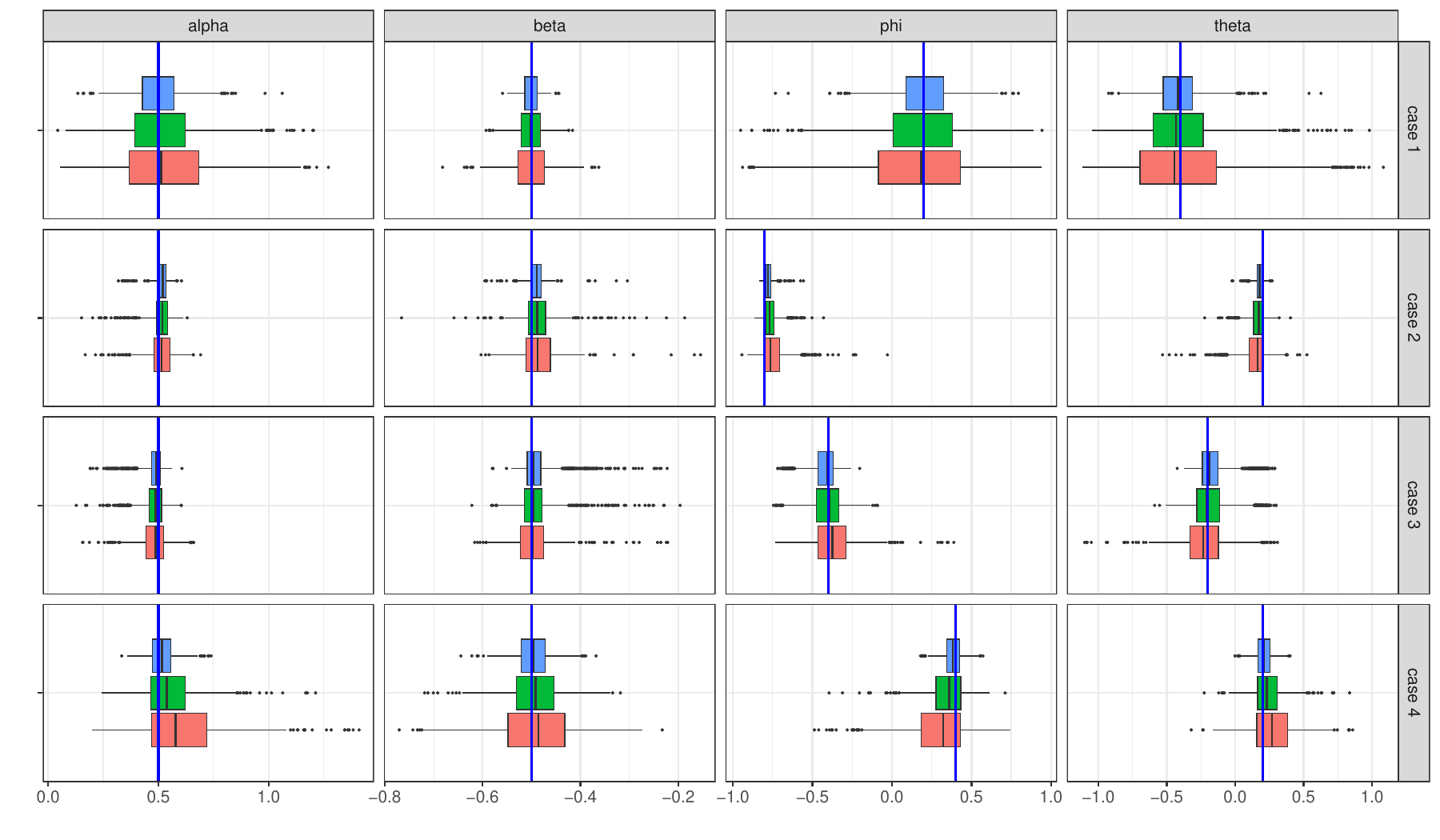}
\includegraphics[width=0.9\textwidth]{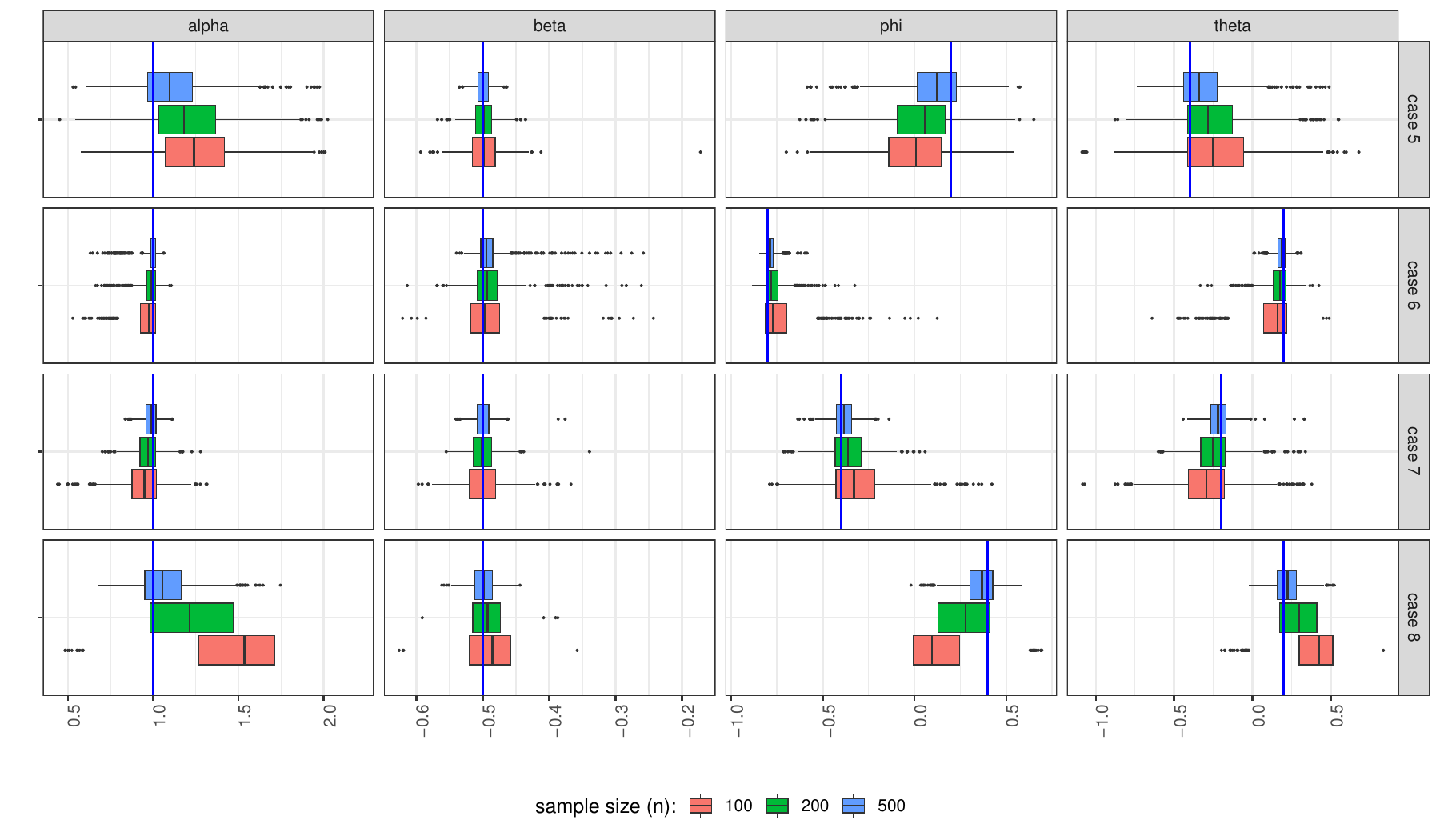}
\caption{Boxplots of the simulation results for all parameter for $\alpha=0.5$ (top) and $\alpha=1$ (bottom), with $\beta=-0.5$ fixed. Parameter $(\phi,\theta)$ are presented as follows. Cases 1 and 5: $(0.2,-0.4)$, cases 2 and 6: $(-0.8,0.2)$, cases 3 and 7: $(-0.4,-0.2)$, and cases 4 and 8: $(0.4,0.2)$. Vertical blue lines indicate the true parameter value}. \label{bp}
\end{figure}

Table \ref{point} summarizes the simulation results. For each set of parameters, we present the mean (left), median (center, in italics), and standard deviations (in parentheses) calculated from the 1,000 replicas. The table shows that parameter $\beta$ is remarkably well estimated in all cases. Parameter $\alpha$ is also well estimated, especially when $n=500$. A comparison between the mean and median estimates shows that for $n=100$, the estimates are slightly skewed. This can be better visualized in the boxplots in Figure \ref{bp}. For most parameter combinations, estimation for $n=500$ is very good, while for $n=100$, it is typically fair. One exception is $(\alpha,\theta,\phi)=(1,0.2,-0.4)$ (case 5 in Figure \ref{bp}), which presents high bias even for $n=500$. Interestingly, $(\alpha,\theta,\phi)=(0.5,0.2,-0.4)$ (case 1 in Figure \ref{bp}) is well estimated. The estimates for $\alpha=1$ for this particular combination were poor because, in this scenario, the peaks of the generated time series get very close to 1, causing the optimization of the log-likelihood to fail to converge. From Table \ref{point} and Figure \ref{bp}, it is perceptible that the variance of the PMLE differs depending on the parameter combination.
\begin{table}[!ht]
\renewcommand{\arraystretch}{1.1}
\setlength{\tabcolsep}{3.6pt}
\caption{Simulation results - point estimates based on 1,000 replicas of each scenario. For each $n$, $\alpha$, $\beta$, $\phi$, and $\theta$, the presented values correspond to the mean (left), the median (center, in italic) and the standard deviation (right, in parentheses).}\label{point}
\centering
\scriptsize
\vspace{.3cm}
\begin{tabular}{c|ccc|ccc|ccc|ccc}
\hline
\multicolumn{1}{c|}{$n$} & \multicolumn{3}{c|}{$\alpha = 0.5$}&\multicolumn{3}{c|}{$\beta = -0.5$} &\multicolumn{3}{c|}{$\phi = 0.2$}&\multicolumn{3}{c}{$\theta = -0.4$}\\
 \hline
100 & 0.539  & \emph{ 0.513 } & $(0.238)$ & -0.500  & \emph{ -0.499 } & $(0.043)$ & 0.145  & \emph{ 0.185 } & $(0.392)$ & -0.391  & \emph{ -0.443 } & $(0.454)$  \\
200 & 0.514  & \emph{ 0.499 } & $(0.184)$ & -0.502  & \emph{ -0.502 } & $(0.029)$ & 0.180  & \emph{ 0.199 } & $(0.300)$ & -0.398  & \emph{ -0.432 } & $(0.310)$  \\
500 & 0.501  & \emph{ 0.500 } & $(0.114)$ & -0.501  & \emph{ -0.501 } & $(0.018)$ & 0.200  & \emph{ 0.201 } & $(0.187)$ & -0.408  & \emph{ -0.417 } & $(0.179)$  \\
\hline
\multicolumn{1}{c|}{$n$} & \multicolumn{3}{c|}{$\alpha = 0.5$}&\multicolumn{3}{c|}{$\beta = -0.5$} &\multicolumn{3}{c|}{$\phi = -0.8$}&\multicolumn{3}{c}{$\theta = 0.2$}\\
 \hline
100 & 0.513  & \emph{ 0.515 } & $(0.061)$ & -0.486  & \emph{ -0.488 } & $(0.042)$ & -0.745  & \emph{ -0.764 } & $(0.087)$ & 0.147  & \emph{ 0.164 } & $(0.108)$  \\
200 & 0.512  & \emph{ 0.518 } & $(0.054)$ & -0.487  & \emph{ -0.489 } & $(0.037)$ & -0.761  & \emph{ -0.770 } & $(0.050)$ & 0.164  & \emph{ 0.172 } & $(0.061)$  \\
500 & 0.518  & \emph{ 0.520 } & $(0.030)$ & -0.490  & \emph{ -0.489 } & $(0.020)$ & -0.774  & \emph{ -0.779 } & $(0.027)$ & 0.177  & \emph{ 0.181 } & $(0.032)$  \\
\hline
\multicolumn{1}{c|}{$n$} & \multicolumn{3}{c|}{$\alpha = 0.5$}&\multicolumn{3}{c|}{$\beta = -0.5$} &\multicolumn{3}{c|}{$\phi = -0.4$}&\multicolumn{3}{c}{$\theta = -0.2$}\\
 \hline
100 & 0.484  & \emph{ 0.488 } & $(0.064)$ & -0.497  & \emph{ -0.498 } & $(0.043)$ & -0.372  & \emph{ -0.376 } & $(0.151)$ & -0.227  & \emph{ -0.234 } & $(0.186)$  \\
200 & 0.481  & \emph{ 0.488 } & $(0.058)$ & -0.490  & \emph{ -0.497 } & $(0.045)$ & -0.410  & \emph{ -0.395 } & $(0.116)$ & -0.173  & \emph{ -0.208 } & $(0.162)$  \\
500 & 0.477  & \emph{ 0.492 } & $(0.056)$ & -0.485  & \emph{ -0.497 } & $(0.044)$ & -0.431  & \emph{ -0.408 } & $(0.095)$ & -0.139  & \emph{ -0.188 } & $(0.157)$  \\
\hline
\multicolumn{1}{c|}{$n$} & \multicolumn{3}{c|}{$\alpha = 0.5$}&\multicolumn{3}{c|}{$\beta = -0.5$} &\multicolumn{3}{c|}{$\phi = 0.4$}&\multicolumn{3}{c}{$\theta = 0.2$}\\
 \hline
100 & 0.609  & \emph{ 0.578 } & $(0.190)$ & -0.490  & \emph{ -0.486 } & $(0.082)$ & 0.296  & \emph{ 0.322 } & $(0.187)$ & 0.271  & \emph{ 0.269 } & $(0.171)$  \\
200 & 0.552  & \emph{ 0.539 } & $(0.119)$ & -0.493  & \emph{ -0.492 } & $(0.058)$ & 0.351  & \emph{ 0.360 } & $(0.118)$ & 0.233  & \emph{ 0.230 } & $(0.116)$  \\
500 & 0.519  & \emph{ 0.517 } & $(0.066)$ & -0.497  & \emph{ -0.497 } & $(0.038)$ & 0.383  & \emph{ 0.382 } & $(0.066)$ & 0.210  & \emph{ 0.211 } & $(0.066)$  \\
\hline
\multicolumn{1}{c|}{$n$} & \multicolumn{3}{c|}{$\alpha = 1$}&\multicolumn{3}{c|}{$\beta = -0.5$} &\multicolumn{3}{c|}{$\phi = 0.2$}&\multicolumn{3}{c}{$\theta = -0.4$}\\
 \hline
100 & 1.252  & \emph{ 1.239 } & $(0.258)$ & -0.498  & \emph{ -0.499 } & $(0.028)$ & -0.001  & \emph{ 0.009 } & $(0.211)$ & -0.237  & \emph{ -0.249 } & $(0.276)$  \\
200 & 1.204  & \emph{ 1.181 } & $(0.250)$ & -0.499  & \emph{ -0.498 } & $(0.018)$ & 0.037  & \emph{ 0.057 } & $(0.202)$ & -0.260  & \emph{ -0.284 } & $(0.229)$  \\
500 & 1.115  & \emph{ 1.097 } & $(0.213)$ & -0.499  & \emph{ -0.499 } & $(0.011)$ & 0.108  & \emph{ 0.124 } & $(0.173)$ & -0.321  & \emph{ -0.342 } & $(0.179)$  \\
\hline
\multicolumn{1}{c|}{$n$} & \multicolumn{3}{c|}{$\alpha = 1$}&\multicolumn{3}{c|}{$\beta = -0.5$} &\multicolumn{3}{c|}{$\phi = -0.8$}&\multicolumn{3}{c}{$\theta = 0.2$}\\
 \hline
100 & 0.963  & \emph{ 0.975 } & $(0.082)$ & -0.494  & \emph{ -0.496 } & $(0.039)$ & -0.737  & \emph{ -0.769 } & $(0.120)$ & 0.130  & \emph{ 0.160 } & $(0.143)$  \\
200 & 0.980  & \emph{ 0.989 } & $(0.060)$ & -0.491  & \emph{ -0.494 } & $(0.032)$ & -0.767  & \emph{ -0.781 } & $(0.060)$ & 0.166  & \emph{ 0.176 } & $(0.078)$  \\
500 & 0.989  & \emph{ 1.000 } & $(0.053)$ & -0.490  & \emph{ -0.494 } & $(0.027)$ & -0.781  & \emph{ -0.787 } & $(0.030)$ & 0.184  & \emph{ 0.186 } & $(0.038)$  \\
\hline
\multicolumn{1}{c|}{$n$} & \multicolumn{3}{c|}{$\alpha = 1$}&\multicolumn{3}{c|}{$\beta = -0.5$} &\multicolumn{3}{c|}{$\phi = -0.4$}&\multicolumn{3}{c}{$\theta = -0.2$}\\
 \hline
100 & 0.943  & \emph{ 0.949 } & $(0.117)$ & -0.500  & \emph{ -0.500 } & $(0.031)$ & -0.319  & \emph{ -0.329 } & $(0.174)$ & -0.292  & \emph{ -0.294 } & $(0.196)$  \\
200 & 0.970  & \emph{ 0.971 } & $(0.074)$ & -0.500  & \emph{ -0.501 } & $(0.021)$ & -0.358  & \emph{ -0.363 } & $(0.111)$ & -0.246  & \emph{ -0.250 } & $(0.130)$  \\
500 & 0.988  & \emph{ 0.988 } & $(0.043)$ & -0.500  & \emph{ -0.500 } & $(0.014)$ & -0.383  & \emph{ -0.384 } & $(0.065)$ & -0.218  & \emph{ -0.220 } & $(0.078)$  \\
\hline
\multicolumn{1}{c|}{$n$} & \multicolumn{3}{c|}{$\alpha = 1$}&\multicolumn{3}{c|}{$\beta = -0.5$} &\multicolumn{3}{c|}{$\phi = 0.4$}&\multicolumn{3}{c}{$\theta = 0.2$}\\
 \hline
100 & 1.451  & \emph{ 1.534 } & $(0.354)$ & -0.488  & \emph{ -0.486 } & $(0.043)$ & 0.144  & \emph{ 0.095 } & $(0.203)$ & 0.387  & \emph{ 0.426 } & $(0.181)$  \\
200 & 1.230  & \emph{ 1.214 } & $(0.302)$ & -0.494  & \emph{ -0.493 } & $(0.030)$ & 0.269  & \emph{ 0.279 } & $(0.173)$ & 0.290  & \emph{ 0.296 } & $(0.155)$  \\
500 & 1.070  & \emph{ 1.055 } & $(0.168)$ & -0.498  & \emph{ -0.498 } & $(0.020)$ & 0.360  & \emph{ 0.368 } & $(0.096)$ & 0.224  & \emph{ 0.224 } & $(0.092)$  \\
\hline
\end{tabular}
\end{table}

\subsubsection*{Simulation results: joint asymptotic behavior}
\begin{figure}[!ht]
\centering
\includegraphics[width=0.48\textwidth]{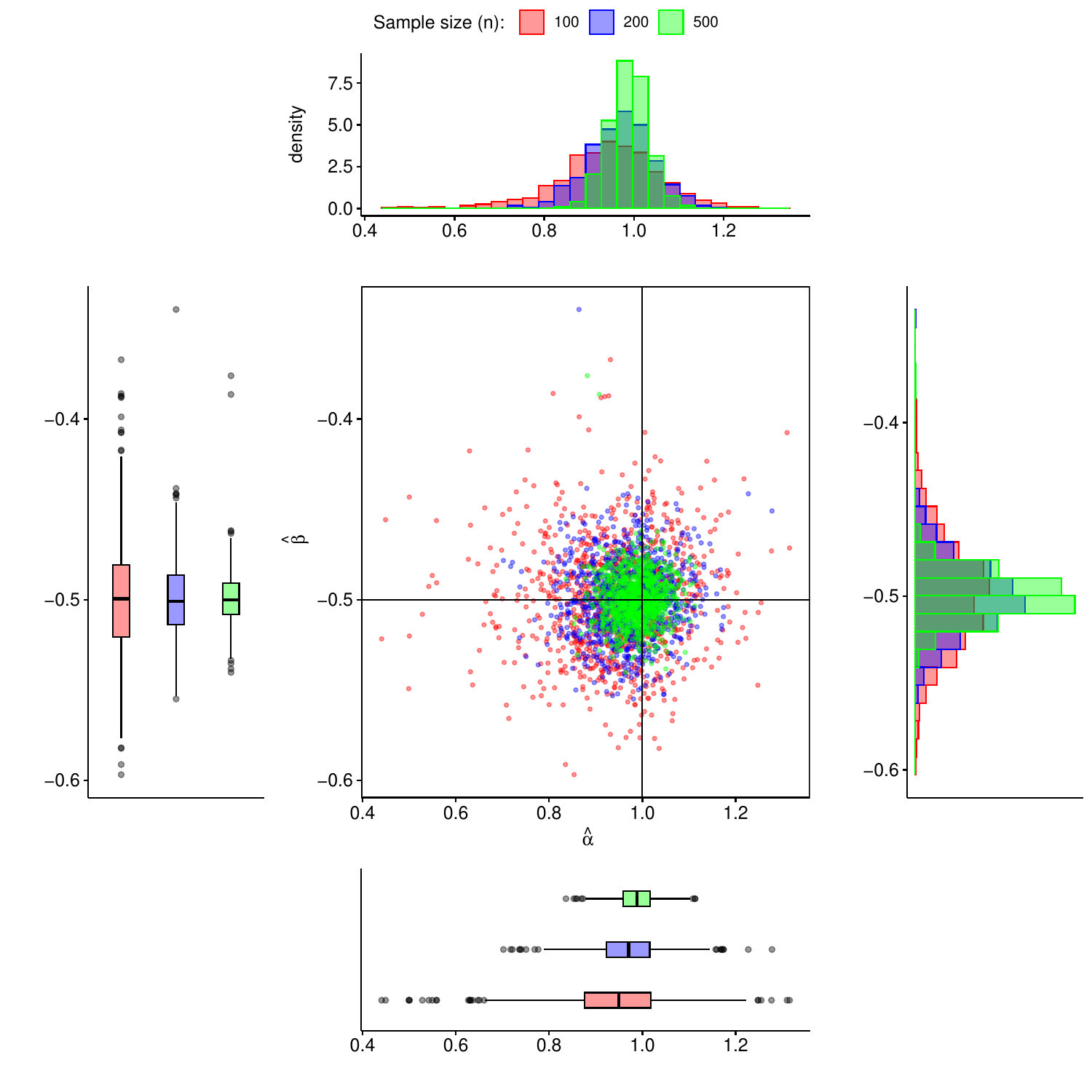}
\includegraphics[width=0.48\textwidth]{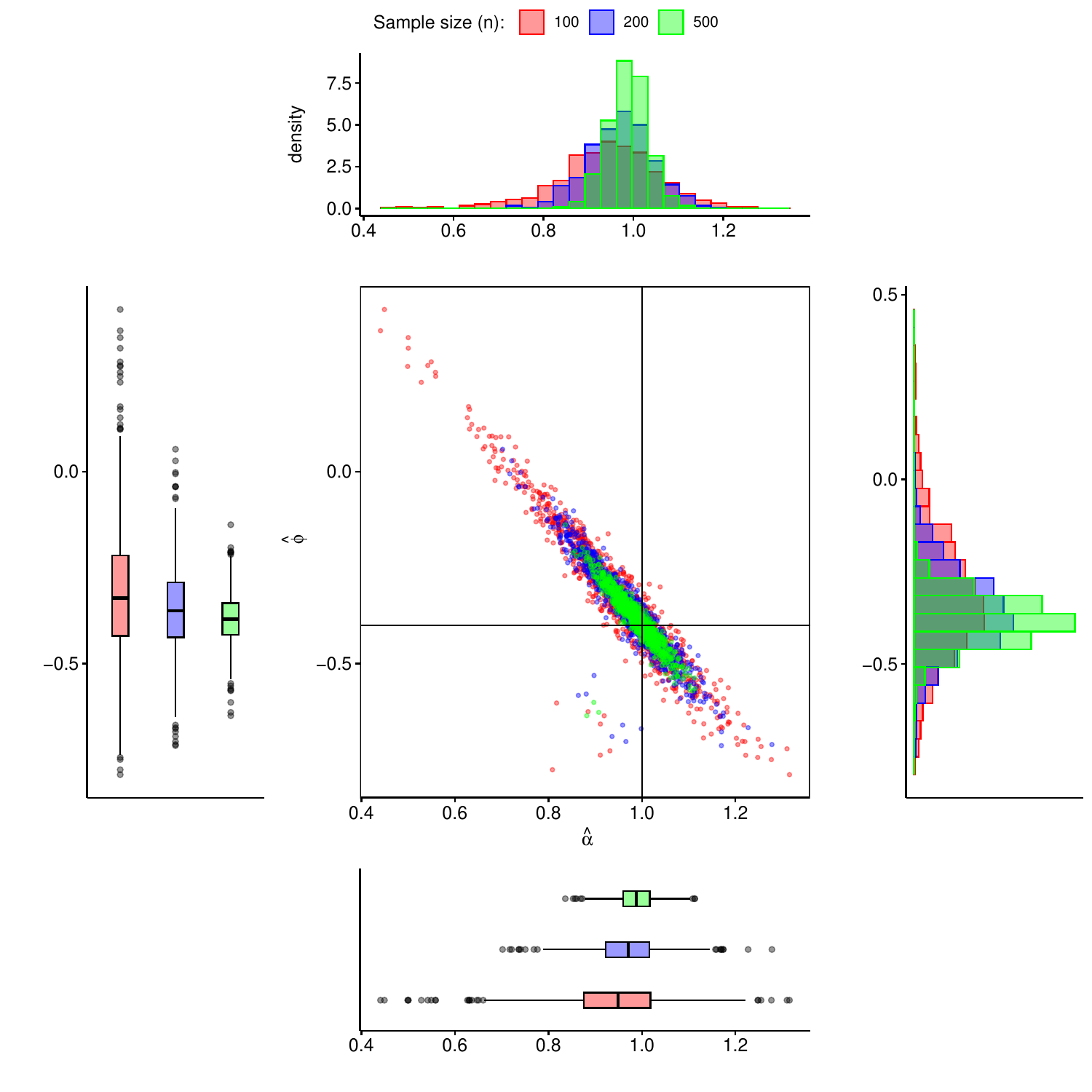}
\includegraphics[width=0.48\textwidth]{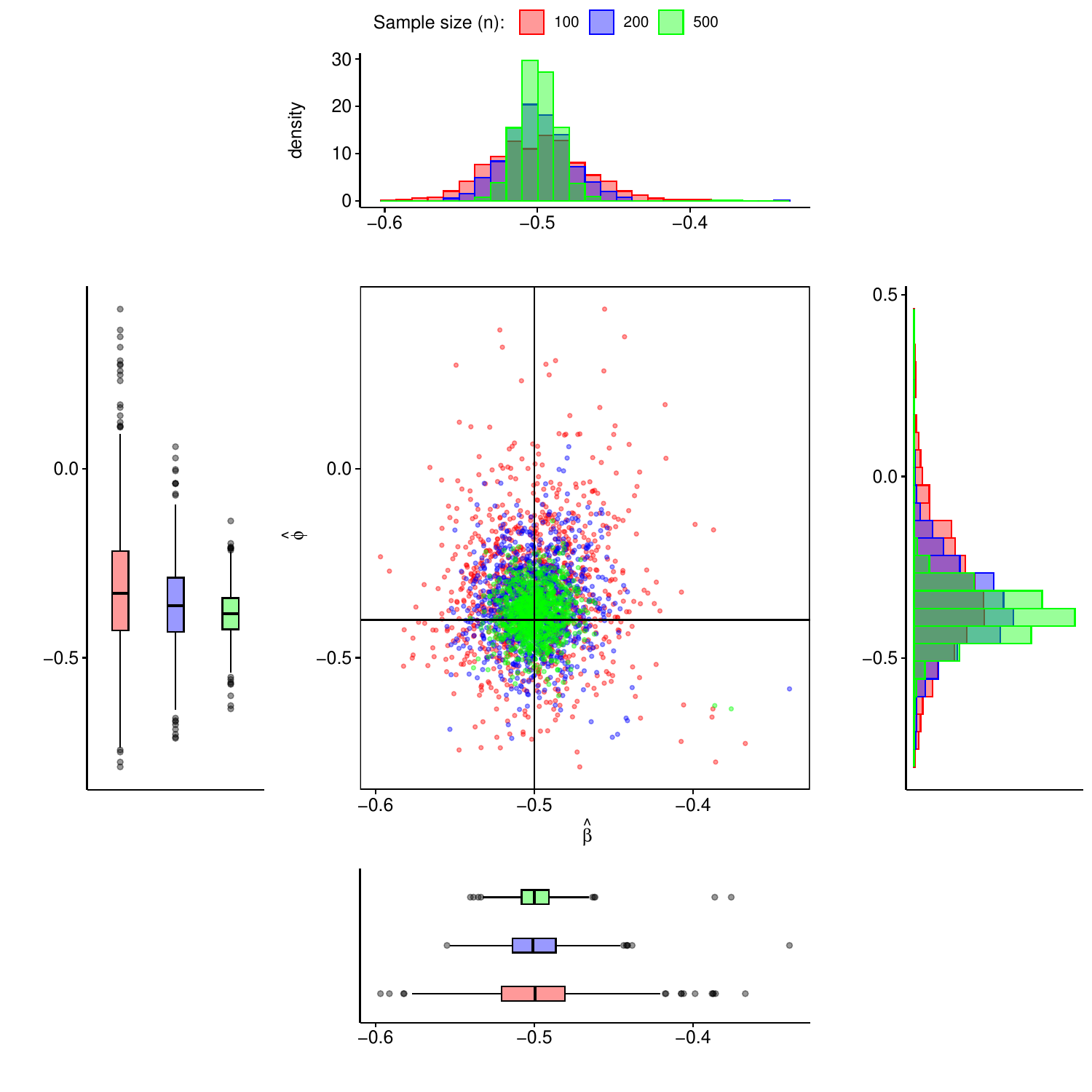}
\includegraphics[width=0.48\textwidth]{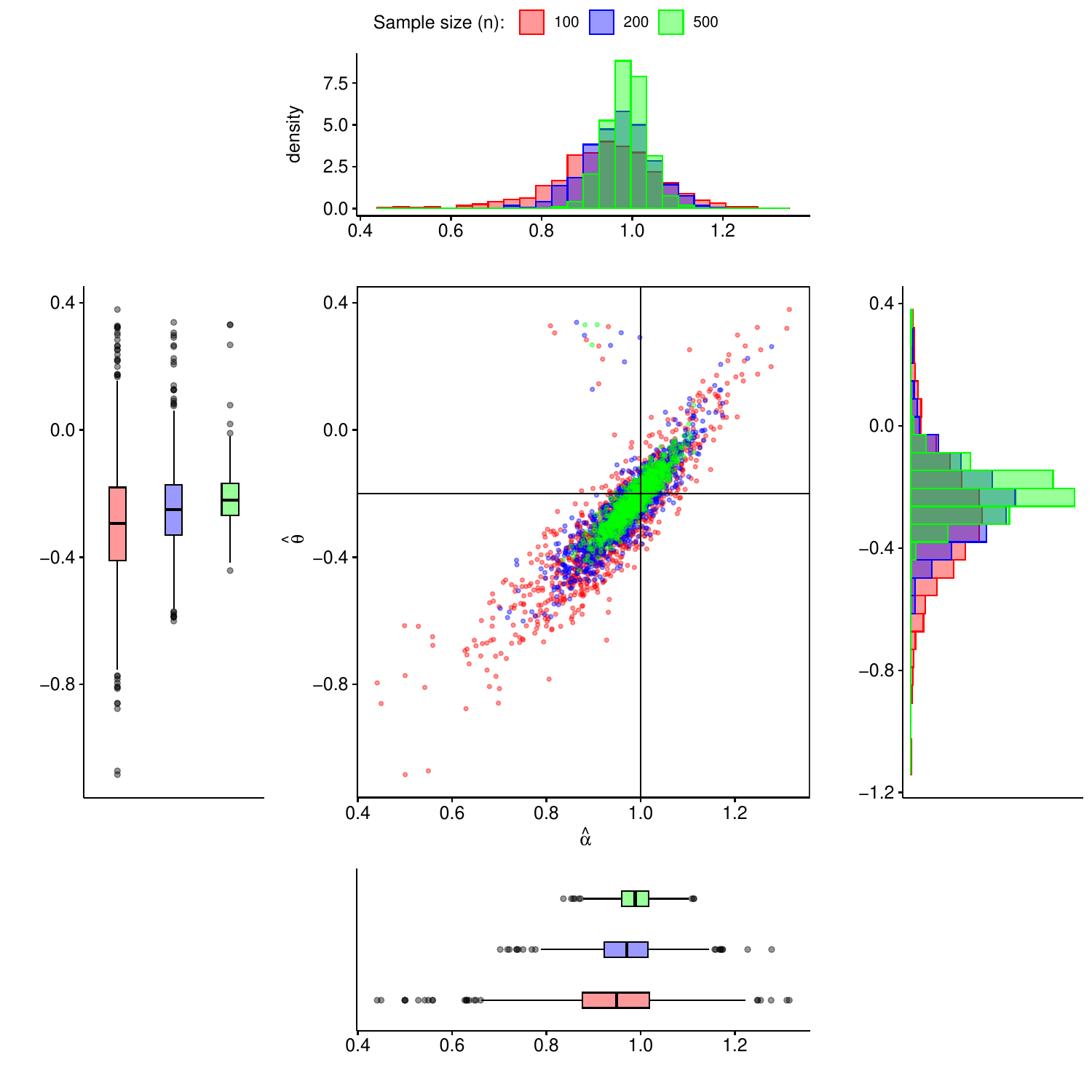}
\includegraphics[width=0.48\textwidth]{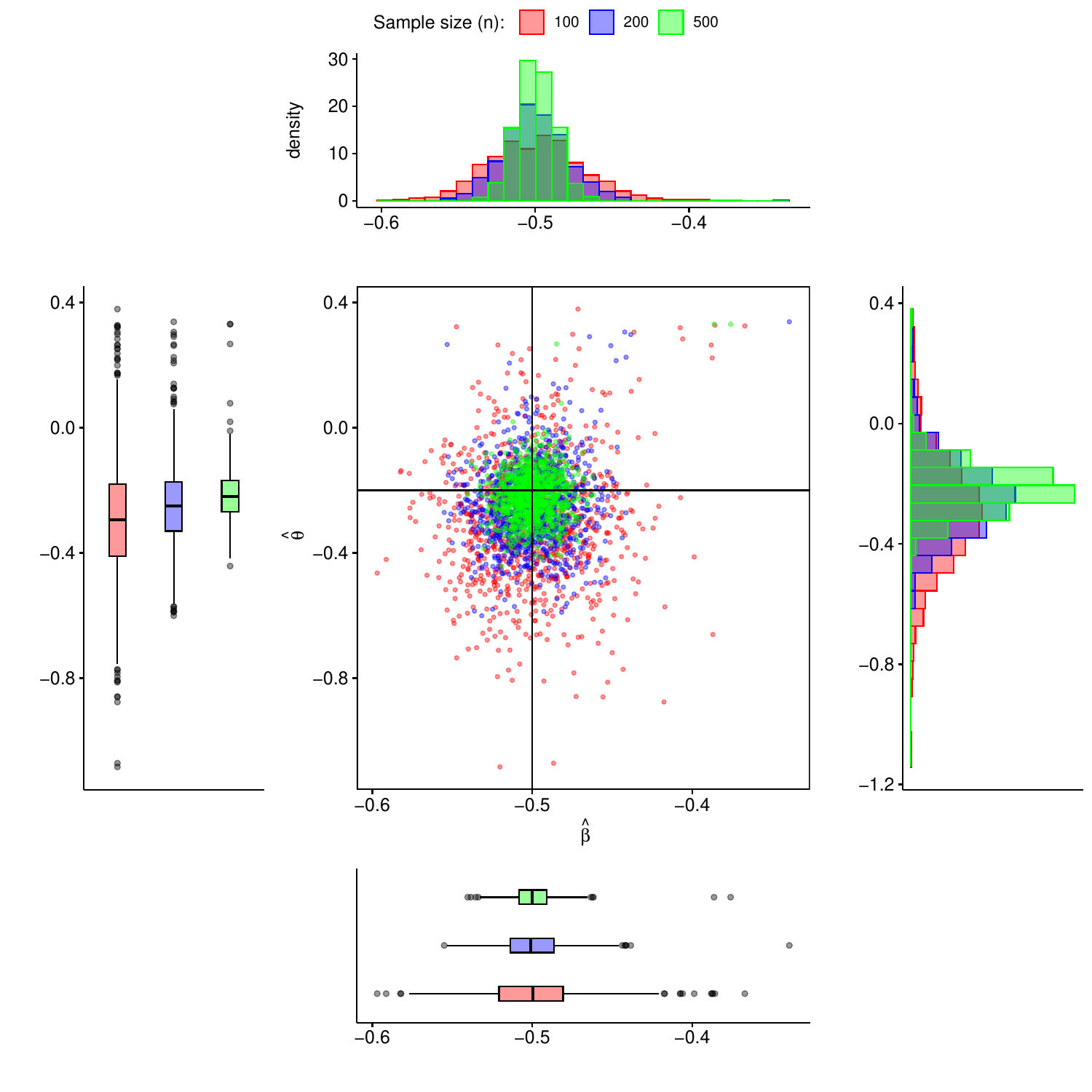}
\includegraphics[width=0.48\textwidth]{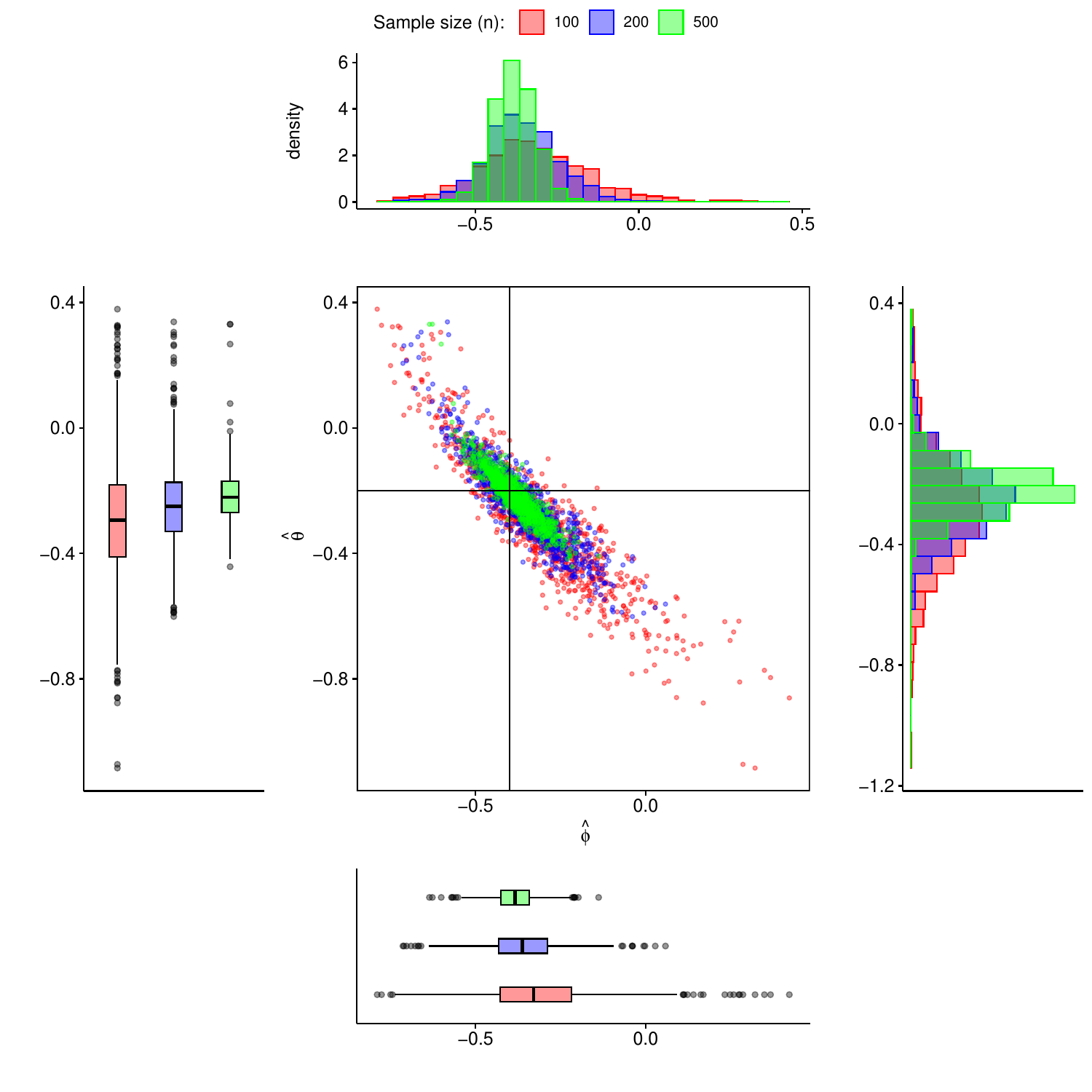}
\caption{Pairwise joint and marginal behavior of the estimated values for $\alpha=1$, $\beta=-0.5$, $\phi=-0.4$, $\theta=-0.2$. Solid lines in the scatter plot represent the true values.}
\label{pair}
\end{figure}

To investigate the asymptotic normality of the PMLE we examine pairwise scatter plots and marginal behavior (histograms and boxplots) presented in Figure \ref{pair} for the case $\alpha=1$, $\beta=-0.5$, $\phi=-0.4$, $\theta=-0.2$. Other cases are presented in the supplementary material and reveal similar behavior. From the scatter plots, we observe the convergence of the points to a familiar Gaussian behavior as $n$ increases and a strong dependence between the estimates of parameters $\alpha$, $\phi$, and $\theta$, presented in the right column of Figure \ref{pair}. In sharp contrast, the estimates of $\beta$ seem uncorrelated with the other parameters, as seen in the plots in the left column of Figure \ref{pair}. The boxplots become more symmetrical as $n$ increases, with the histograms resembling the shape of a normal distribution.
\FloatBarrier
\subsection{Goodness-of-fit tests}

This section examines the finite-sample performance of goodness-of-fit tests based on the simple and quantile residuals discussed in Section \ref{gof}. For the simple residuals, $\hat e_t = Y_t - \hat\mu_t$, where $\hat\mu_t$ is obtained using the estimated parameter values. If the model is well specified, $\hat e_t$ should approximately behave as a martingale difference with respect to the process's history. This can be tested using any martingale difference test, such as the wild bootstrap automatic variance ratio test (WB for short) proposed by \cite{Choi}, adopted here. In our experience, we found that the WB presents a good balance between power and computational speed. The finite-sample performance of this and other methods is discussed in \cite{charles}. The method is implemented in package \texttt{vrtest} \citep{vrtest}.  As for the quantile residuals, under the correct model specification, it should follow a standard normal distribution. For testing purposes, we apply five commonly used normality tests: the Anderson-Darling (AD), Cram\'er-von Mises (CvM), Kolmogorov-Smirnov (KS), implemented in the R package \texttt{nortest} \citep{nortest}, and the Shapiro-Francia (SF), available from base R. Further details on these tests can be found in \cite{thode}.
\subsubsection*{Data generating process}
For this exercise, we generate samples of size $n\in\{100,200,500\}$ from an MARMA$(1,1)$ model with parameters $\alpha\in\{0.5,1\}$, $(\phi,\theta) \in\big\{(0.2,-0.8),(-0.8,0.2),(-0.4,-0.2),(0.4,0.2)\big\}$. No covariates are considered, and $g(\cdot)$ is taken to be the cloglog function. Under these specifications, the underlying model is given by
\begin{equation*}
\eta_t := g(\mu_t)= \alpha +  \phi g(Y_{t-1}) + \theta r_{t-1}, \quad g(x) = \log(-\log(1-x)).
\end{equation*}
A size 100 burn-in is applied to generate the time series, a total of 1,000 replicas of each scenario were generated, and tests were evaluated at a 5\% confidence level. To perform the WB test, we consider 500 bootstrap samples using the Mammen's two point distribution (see package \texttt{vrtest}'s documentation for details).
\subsubsection*{Simulation results}
Table \ref{tests} summarizes the simulation results. All tests performed well, with a rejection rate close to the nominal value of 0.05 for most parameters. The normality tests presented overall similar results, which is to be expected. The only noticeable exception is the case $\alpha=1$, $\phi=-0.8$, and $\theta=0.2$, where the SF test presented somewhat higher-than-expected rejection rates.
\begin{table}[!ht]
\renewcommand{\arraystretch}{1.2}
\setlength{\tabcolsep}{5pt}
\caption{Simulation results - martingale difference, and normality tests. For each $n$,$\alpha$, and $(\phi, \theta)$, the results presented correspond to the proportion of tests that rejected the null hypothesis for each specific test.}\label{tests}
\centering
\footnotesize
\vspace{.3cm}
\begin{tabular}{c|c|c||c|c|c|c||c||c|c|c|c}
\hline
 \hline
  \multirow{2}{*}{$n$} &\multirow{2}{*}{$(\phi,\theta)$}&\multicolumn{5}{c||}{$\alpha = 0.5$}&\multicolumn{5}{c}{$\alpha = 1$}\\
  \cline{3-12}
  & & WB &  AD & CvM & KS & SF & WB &  AD & CvM & KS & SF \\
  \hline
100 & \multirow{3}{1.5cm}{\begin{tabular}{l}
        $\phi = 0.2$\\ $\theta = -0.4$ \\
        \end{tabular}}
 & 0.03 &  0.06 &  0.06 &  0.05 &  0.06 & 0.03 &  0.04 &  0.05 &  0.05 & 0.04 \\
200 &  & 0.01 &  0.05 &  0.05 &  0.06 &  0.05 & 0.02 &  0.06 &  0.06 &  0.05 & 0.06 \\
500 &  & 0.00 &  0.05 &  0.06 &  0.05 &  0.06 & 0.01 &  0.05 &  0.05 &  0.05 & 0.04 \\
\hline
100 & \multirow{3}{1.5cm}{\begin{tabular}{l}
        $\phi = -0.8$\\ $\theta = 0.2$ \\
        \end{tabular}}
 & 0.04 &  0.06 &  0.05 &  0.06 &  0.08 & 0.06 &  0.09 &  0.08 &  0.06 & 0.17 \\
200 &  & 0.06 &  0.06 &  0.05 &  0.05 &  0.09 & 0.06 &  0.07 &  0.07 &  0.06 & 0.15 \\
500 &  & 0.06 &  0.04 &  0.04 &  0.03 &  0.05 & 0.07 &  0.05 &  0.05 &  0.04 & 0.12 \\
\hline
100 & \multirow{3}{1.5cm}{\begin{tabular}{l}
        $\phi = -0.4$\\ $\theta = -0.2$ \\
        \end{tabular}}
 & 0.03 &  0.05 &  0.05 &  0.05 &  0.06 & 0.02 &  0.05 &  0.05 &  0.06 & 0.06 \\
200 &  & 0.05 &  0.06 &  0.06 &  0.06 &  0.07 & 0.03 &  0.06 &  0.06 &  0.06 & 0.08 \\
500 &  & 0.09 &  0.05 &  0.05 &  0.04 &  0.10 & 0.03 &  0.04 &  0.04 &  0.05 & 0.08 \\
\hline
100 & \multirow{3}{1.5cm}{\begin{tabular}{l}
        $\phi = 0.4$\\ $\theta = 0.2$ \\
        \end{tabular}}
 & 0.04 &  0.05 &  0.05 &  0.06 &  0.05 & 0.04 &  0.07 &  0.07 &  0.06 & 0.06 \\
200 &  & 0.03 &  0.05 &  0.06 &  0.06 &  0.05 & 0.05 &  0.07 &  0.07 &  0.06 & 0.06 \\
500 &  & 0.04 &  0.05 &  0.05 &  0.05 &  0.05 & 0.07 &  0.06 &  0.06 &  0.05 & 0.06 \\
\hline
\end{tabular}
\end{table}

\subsection{Prediction confidence interval}
In this exercise, we explore the finite-sample performance of the bootstrap approach for constructing prediction confidence intervals discussed in Section \ref{msf}.

\subsubsection*{Data generating process and intervals construction}
For this study, for simplicity, we fix $\alpha = 1$. The remaining parameters follow the same settings as in the goodness-of-fit exercise: we generate MARMA$(1,1)$ models with parameters $(\phi,\theta) \in\big\{(0.2,-0.8),(-0.8,0.2),(-0.4,-0.2),(0.4,0.2)\big\}$, no covariates, cloglog as the link function, and samples of size $n\in\{100,200,500\}$.   Under these specifications, the underlying model is given by
\begin{equation*}
\eta_t := g(\mu_t)= 1 +  \phi g(Y_{t-1}) + \theta r_{t-1}, \quad g(x) = \log(-\log(1-x)).
\end{equation*}
A size 100 burn-in was applied to generate the time series, and a total of 1,000 replicas were considered.

For each time series, we estimate the parameter $\bs\gamma=(\alpha,\phi,\theta)$ via PMLE and use this estimate to create $m=500$ bootstrap samples, as explained in Section \ref{msf}, considering $h=50$ as forecasting horizon. Confidence intervals of level $\delta\in\{0.1,0.05,0.01\}$ are obtained as explained in Section \ref{msf}. The R package \texttt{BTSR} was used to obtain the bootstrap samples, while the R function \texttt{quantile} was used to obtain the lower and upper boundaries of the confidence intervals.

\subsubsection*{Simulation Results}
\begin{figure}[!ht]
\centering
\includegraphics[width=\textwidth]{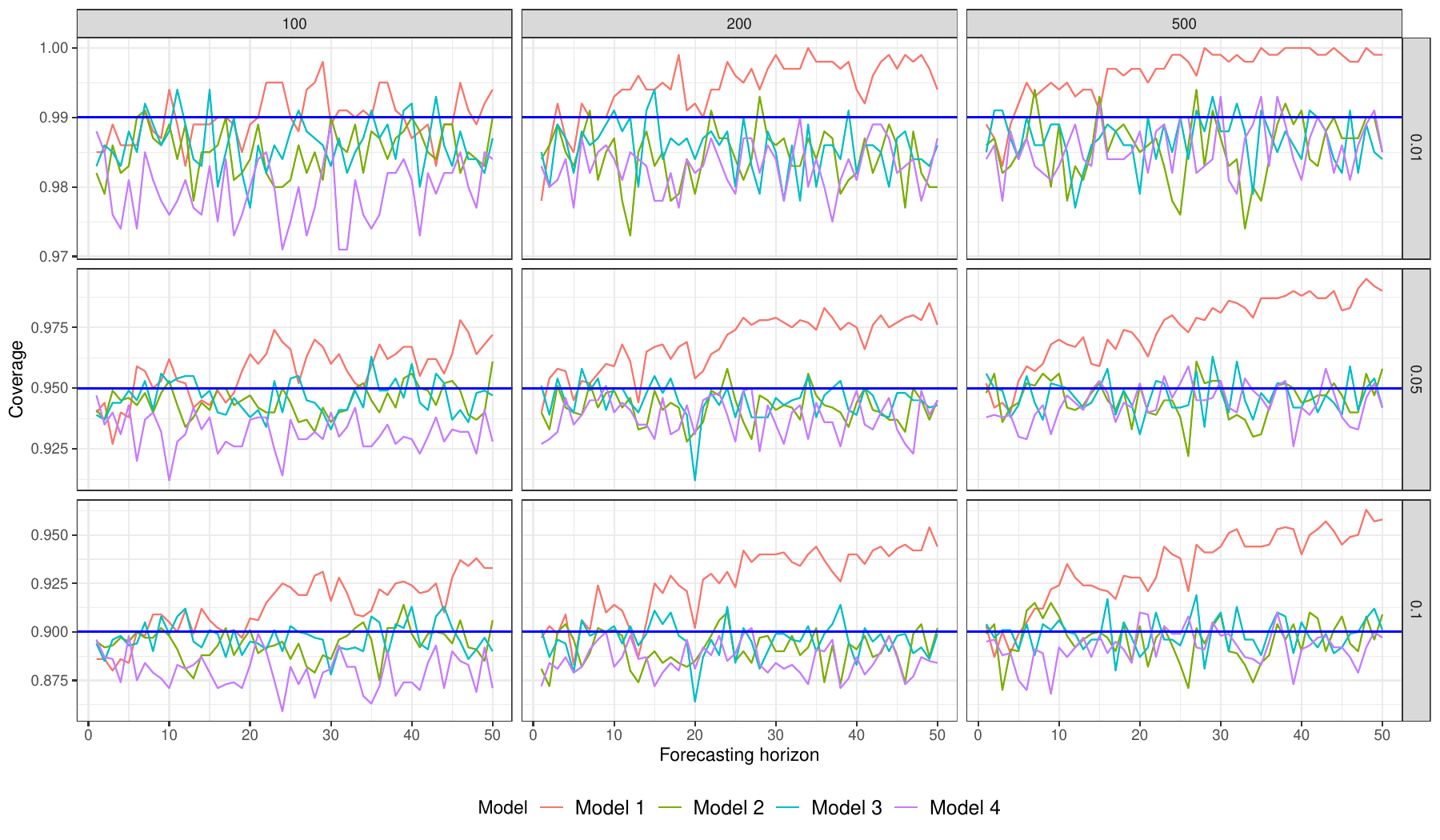}
\caption{Simulation results for the prediction confidence interval exercise. Each block presents the coverage of the bootstrap prediction confidence interval for a given $n$ (column) and $\delta$ (row) as a function of the forecasting horizon $h$. Model 1 refers to parameter $\phi=-0.8$ and $\theta=0.2$; Model 2: $\phi=-0.4$ and $\theta=-0.2$; Model 3: $\phi=0.2$ and $\theta=-0.4$; Model 4: $\phi=0.2$ and $\theta=0.4$. The nominal level is indicated by the blue lines.} \label{predg}
\end{figure}

Figure \ref{predg} presents the simulation results. Each plot presents the coverage of the bootstrap prediction confidence interval for a given $n$ (column) and $\delta$ (row) as a function of the forecasting horizon $h$ and each model. For simplicity, in the legend model 1 refers to parameters $\phi=-0.8$ and $\theta=0.2$, model 2 represents $\phi=-0.4$ and $\theta=-0.2$, model 3 represents $\phi=0.2$ and $\theta=-0.4$, while model 4 represents $\phi=0.2$ and $\theta=0.4$. The plots show that the results are very good overall, with coverage close to the nominal levels for all forecasting lags, even when $n=100$, except perhaps for Model 1. In that particular case, the bootstrap confidence intervals are somewhat conservatory and probably wider than ideal.
\FloatBarrier

\section{Application to real data}\label{sec:Aplication}

This section evaluates the performance of the proposed model in forecasting the monthly useful water volume (UV) of the Guarapiranga Reservoir, located at the border between Itapecerica da Serra and Embu-Gua\c cu, SP, Brazil. Accurately modeling the useful water volume of a reservoir is crucial for effective water resource management. This estimation is key to informed decision-making, and for water management plans to be successful, they must rely on reliable and accurate models~\citep{dang2020}. 

\subsubsection*{Data description}
\begin{figure}[!ht]
\centering
\subfigure[Observed data]{\includegraphics[width=0.49\textwidth]{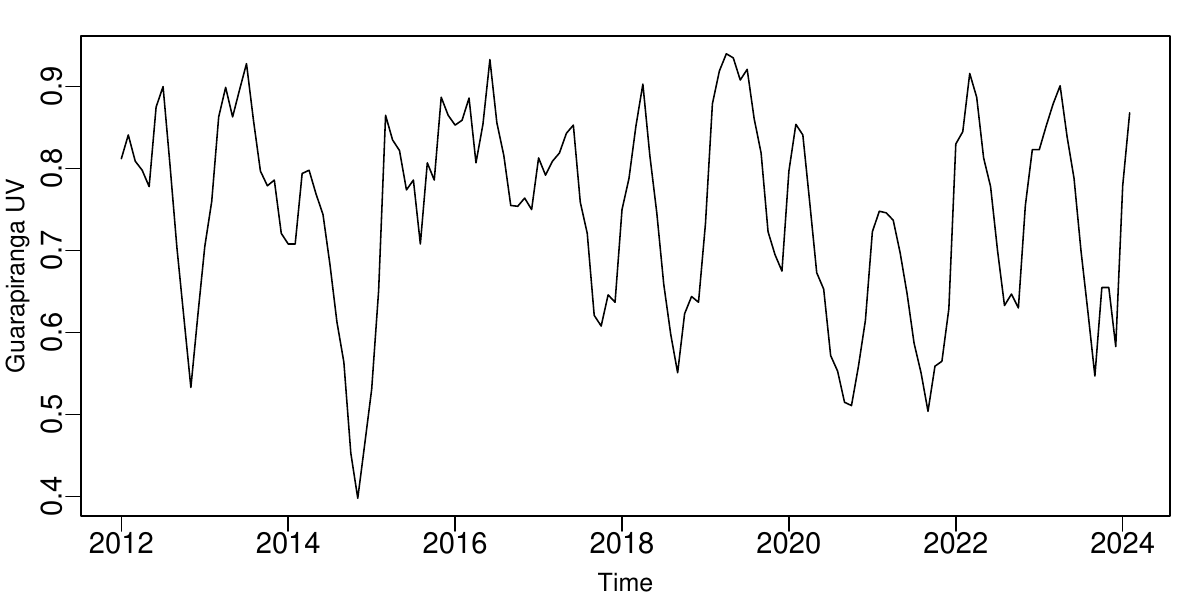}\label{f:ts}}
\subfigure[Sample ACF]{\includegraphics[width=0.49\textwidth]{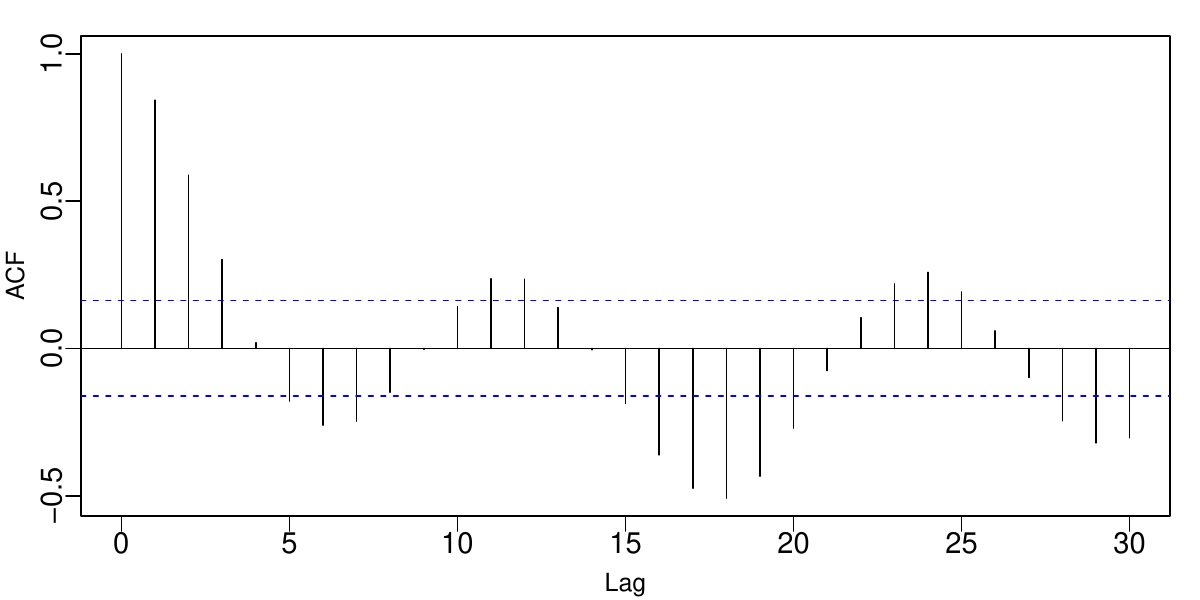}\label{f:acf}}
\subfigure[Sample PACF]{\includegraphics[width=0.49\textwidth]{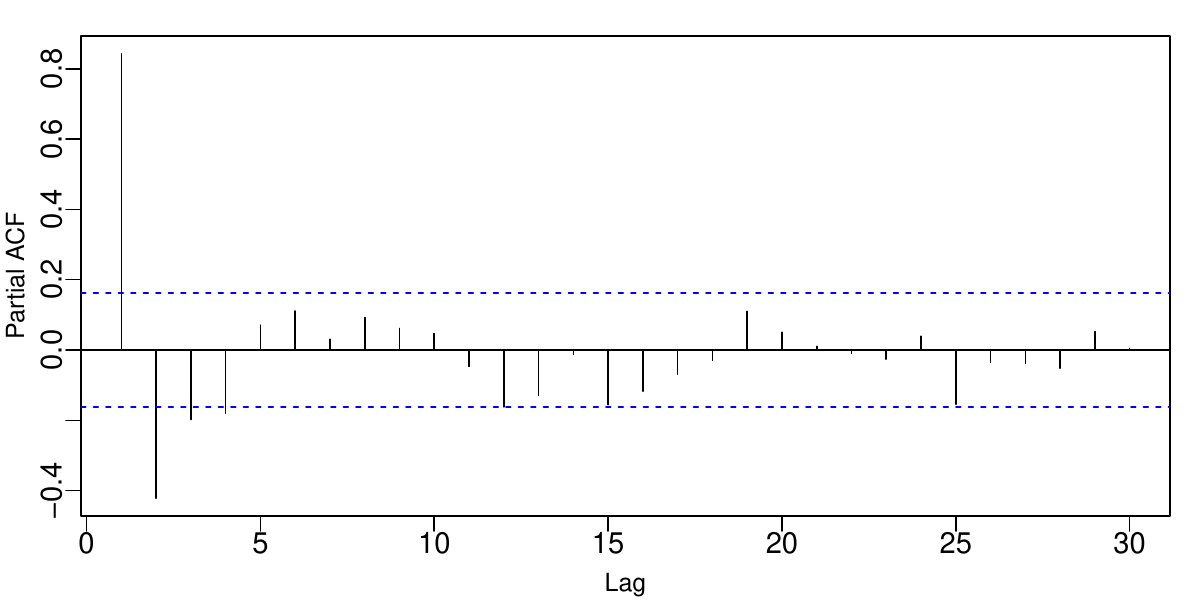}\label{f:pacf}}
\subfigure[Seasonality]{\includegraphics[width=0.49\textwidth]{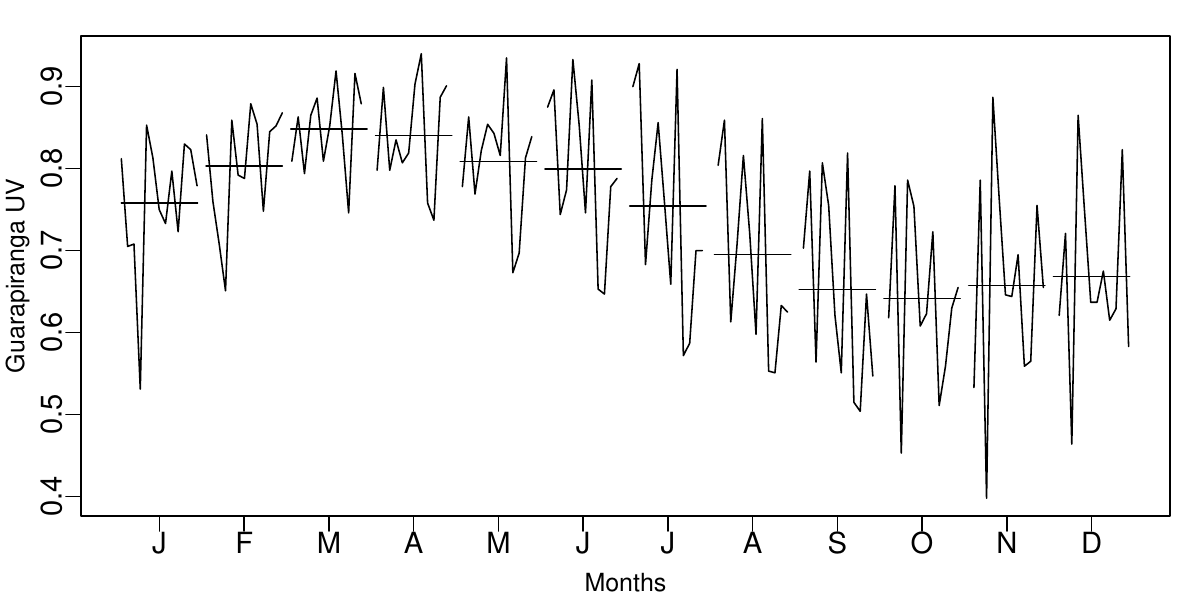}\label{f:sazo}}
\caption{Guarapiranga UV data, Brazil: (a) the observed time series and the corresponding (b) autocorrelation function (ACF), (c) partial autocorrelation (PACF) function and (d) seasonal plot.}\label{fig:f}
\end{figure}

The dataset used for this analysis spans from January 2012 to February 2024, resulting in a sample size of $n = 146$. The observed time series is presented in Figure~\ref{f:ts}. The values range from 0.398 to 0.94, with a median of 0.7665, a sample mean of 0.7444, and a standard deviation of 0.1199. Figures~\ref{f:acf} and~\ref{f:pacf} display the sample autocorrelation function (ACF) and sample partial autocorrelation function (PACF), respectively, while Figure~\ref{f:sazo} shows a seasonal plot. These figures suggest the presence of seasonality with an annual frequency, which aligns with expectations, as reservoirs in that region typically reach higher useful water volume levels from December to April, following the rainy season.

\subsubsection*{Model identification and selection criteria}
The dataset exhibits clear evidence of strong annual seasonality, as indicated by the ACF, PACF, and seasonal plot in Figure~\ref{fig:f}. We consider a MARMA$(p,q)$ model with harmonic regression terms as covariates to capture this periodic pattern. Specifically, we define the covariate vector as
\begin{equation}
\bs{X}_t = \biggl(\sin\Bigl(\frac{2\pi t}{12}\Bigr),\, \cos\Bigl(\frac{2\pi t}{12}\Bigr)\biggr)', \quad t \in \{1, \dots, n\},\label{eq:Xt}
\end{equation}
which allows the model to account for the observed 12-month seasonality. Including these harmonic terms improves the model's ability to capture annual reservoir-level fluctuations, enhancing estimation and forecasting accuracy.

The initial $n = 134$ observations were used for model estimation and selection. The most recent $h = 12$ observations were reserved for forecasting evaluation. Parameter estimation was performed using the L-BFGS-B and Nelder-Mead algorithms combined, as implemented in the \texttt{BTSR} package. We tested the logit, loglog, and cloglog link functions to determine the most suitable transformation. The model selection followed a bidirectional stepwise procedure. The process began with a full MARMA(20,20) model, including the covariates $\mathbf{X}_t$ as defined in \eqref{eq:Xt}. During the backward elimination phase, parameters with $p$-values exceeding 15\% were sequentially fixed at zero, ensuring that only statistically significant terms were retained. In the forward selection phase, previously excluded parameters were reassessed and reintroduced if their $p$-values fell below the 10\% significance threshold. This iterative process, guided by the Wald test described in Section~\ref{s:hp}, continued until no further improvements were identified. The procedure enhanced model interpretability by dynamically adjusting the set of fixed parameters while preserving critical autoregressive, moving average, and covariate effects.

To evaluate the out-of-sample forecasting performance, we compared the proposed MARMA model with the KARMA model~\citep{Bayers} and the Holt-Winters additive method~\citep{holt2004forecasting,winters1960forecasting}. The KARMA model is widely applied in hydrological modeling, while the Holt-Winters is a traditional nonparametric technique for time series forecasting. Parameter estimation for the KARMA model followed the same procedure as for the MARMA model since it is also implemented in the \texttt{BTSR} package. The Holt-Winters method is part of the base R, and it was applied considering default configurations. To have a meaningful comparison of the predicted values among the competing models, we computed some goodness-of-fit measures, such as root mean squared error~(RMSE), mean absolute percentage error~(MAPE), and mean directional accuracy~(MDA).  Lower values of RMSE and MAPE indicate better model performance, while a higher MDA reflects improved alignment of the predictions with the directional movements of the time series (upward or downward).

\subsubsection*{Parameters estimation results}

Table~\ref{tab:results} presents a comparative summary of the best MARMA and KARMA models for each link function. The results include the final model order ($p$ and $q$), the number of significant parameters ($k$), the number of function evaluations ($N$) required for convergence, $p$-values from the DL and Shapiro tests, and key in-sample goodness-of-fit measures (RMSE, MAPE, MDA, AIC, BIC, and HQC). The \textbf{bold} values in the table highlight the most optimal results for each model, while the \textit{italicized} values emphasize potential issues with model specification.

\begin{table}[ht]
\centering
\caption{Summary results for the best model related to each link function: order ($p$ and $q$), number of significant parameters ($k$), number of function evaluations ($N$), $p$-values from DL and Shapiro tests, and in-sample goodness-of-fit measures. For each model, \textbf{bold} highlights the most optimal results, while \textit{italics} emphasizes potential issues with model specification.}\vskip 0.5\baselineskip
\label{tab:results}
\setlength{\tabcolsep}{2.6pt}
\renewcommand{\arraystretch}{1.2}
\begin{tabular}{lcccccccccccccccc}
\cline{2-14}
& Link & $p$ & $q$ & $k$ & $N$ &  DL & ST & RMSE & MAPE & MDA & AIC & BIC & HQC \\
\hline
 \multirow{3}{*}{\rotatebox[origin=c]{90}{\footnotesize MARMA}} & logit & 19 & 0 & 4 & 324 &  0.2313 & 0.1310 & 0.0608 & 0.0726 & \textbf{0.6241} &  -210.54 & -198.95 & -212.19 \\
 & loglog & 16 & 17 & 6 & 348 & 0.1418 & 0.1200 & \textbf{0.0587} & 0.0702 & 0.6015 &  -207.64 & -190.26 & -210.11 \\
 & cloglog & 1 & 0 & 3 & 59 &  0.2164 & 0.1143 & 0.0607 & 0.0729 & 0.6015 &  \textbf{-212.55} & \textbf{-203.86} & \textbf{-213.78} \\
 \hline
\multirow{3}{*}{\rotatebox[origin=c]{90}{\footnotesize KARMA}} & logit & 19 & 0 & 8 & 939 &  0.3284 & \textit{0.0016} & \textbf{0.0498} & 0.0552 & 0.6917 & \textbf{-393.64} & \textbf{-370.46} & \textbf{-396.93} \\
 & loglog & 19 & 0 & 8 & 1171 &  0.2537 & \textit{0.0082} & 0.0508 & 0.0562 & 0.7143 &  -388.14 & -364.96 & -391.43 \\
 & cloglog & 19 & 0 & 11 & 408 &  0.5299 & 0.1457 & 0.0568 & 0.0632 & \textbf{0.7293} &  -362.87 & -330.99 & -367.40 \\
\hline
\end{tabular}
\end{table}

Among the MARMA models, the cloglog link function achieved the lowest AIC/BIC/HQC values, indicating superior model fit compared to the logit and loglog alternatives. However, the loglog model yielded the smallest RMSE (0.0587), while the logit model exhibited the highest MDA (0.6241). The DL and Shapiro test results suggest reasonable residual adequacy across all MARMA models, with $p$-values exceeding 0.11 for both tests.  In contrast, the KARMA models consistently outperformed MARMA in terms of RMSE and MAPE, with the logit link function achieving the lowest RMSE (0.0498) and MAPE (0.0552), while the cloglog model exhibited the highest MDA (0.7293). However, the Shapiro test results for the KARMA models indicated deviations from normality for the logit and loglog link functions, which suggests model misspecification.

Model parsimony is a crucial consideration in selecting an appropriate forecasting model, as it balances goodness of fit with complexity to enhance interpretability and generalizability. Table~\ref{tab:results} shows that while the KARMA models achieved lower RMSE and MAPE values compared to MARMA models, they also required a significantly higher number of function evaluations ($N$) to reach convergence (e.g., 1171 for KARMA loglog) and retained more parameters ($k$). In contrast, the MARMA models demonstrated competitive performance with fewer parameters and lower computational cost.  The coefficients for each final MARMA and KARMA model are reported in Table~\ref{t:cacondefit}.

\begin{table}[!ht]
\centering
\caption{Fitted MARMA and KARMA models for the Guarapiranga UV dataset: estimated coefficients and the respective standard errors (in parenthesis).}\label{t:cacondefit}\vskip 0.5\baselineskip
\renewcommand{\arraystretch}{1.1}
\small
\begin{tabular}{ccccccccc}
\hline
\multicolumn{1}{c}{Model} & & \multicolumn{3}{c}{MARMA} & & \multicolumn{3}{c}{KARMA}\\
\multicolumn{1}{c}{Link} & & logit & loglog & cloglog & & logit & loglog & cloglog\\
\cline{1-1} \cline{3-5} \cline{7-9}
Parameter & & Estimate & Estimate & Estimate & & Estimate & Estimate & Estimate \\
\hline
$\alpha$ & & $\begin{array}{c} 0.5735\\ (0.1610) \end{array}$ & $\begin{array}{c} -0.95796\\ (0.1564) \end{array}$ & $\begin{array}{c} 0.1534\\ (0.0472) \end{array}$ & & $\begin{array}{c} 0.7142\\ (0.0787) \end{array}$ & $\begin{array}{c} -0.86857\\ (0.0801) \end{array}$ & $\begin{array}{c} 0.1834\\ (0.0257) \end{array}$ \\
$\beta_1$ & & $\begin{array}{c} 0.4900\\ (0.1522) \end{array}$ & $\begin{array}{c} -0.3992\\ (0.1711) \end{array}$ & $\begin{array}{c} 0.2119\\ (0.0721) \end{array}$ & & $\begin{array}{c} 0.6833\\ (0.2356) \end{array}$ & $\begin{array}{c} -0.5746\\ (0.1759) \end{array}$ & $\begin{array}{c} 3.3778\\ (0.9085) \end{array}$ \\
$\beta_{2}$ & & - & - & - & & - & - & $\begin{array}{c} -1.2221\\ (0.5109) \end{array}$ \\
$\phi_1$ & & $\begin{array}{c} 0.6920\\ (0.0931) \end{array}$ & $\begin{array}{c} 0.6544\\ (0.0711) \end{array}$ & $\begin{array}{c} 0.7152\\ (0.0899) \end{array}$ & & $\begin{array}{c} 0.9895\\ (0.0629) \end{array}$ & $\begin{array}{c} 0.9173\\ (0.0638) \end{array}$ & $\begin{array}{c} 1.2937\\ (0.0821) \end{array}$ \\
$\phi_{2}$ & & - & - & - & & - & - & $\begin{array}{c} -0.2312\\ (0.1317) \end{array}$ \\
$\phi_{3}$ & & - & - & - & & $\begin{array}{c} -0.3070\\ (0.0510) \end{array}$ & $\begin{array}{c} -0.2777\\ (0.0496) \end{array}$ & $\begin{array}{c} -0.4487\\ (0.0893) \end{array}$ \\
$\phi_{5}$ & & - & - & - & & - & - & $\begin{array}{c} 0.0631\\ (0.0390) \end{array}$ \\
$\phi_{11}$ & & - & - & - & & $\begin{array}{c} 0.1057\\ (0.0461) \end{array}$ & - & - \\
$\phi_{12}$ & & - & - & - & & - & $\begin{array}{c} 0.1717\\ (0.0663) \end{array}$ & $\begin{array}{c} 0.0888\\ (0.0429) \end{array}$ \\
$\phi_{13}$ & & - & - & - & & $\begin{array}{c} -0.2278\\ (0.0431) \end{array}$ & $\begin{array}{c} -0.2856\\ (0.0611) \end{array}$ & $\begin{array}{c} -0.1618\\ (0.0438) \end{array}$ \\
$\phi_{16}$ & & - & $\begin{array}{c} -0.3120\\ (0.0717) \end{array}$ & - & & - & - & - \\
$\phi_{19}$ & & $\begin{array}{c} -0.1132\\ (0.0764) \end{array}$ & - & - & & $\begin{array}{c} -0.1519\\ (0.0344) \end{array}$ & $\begin{array}{c} -0.1769\\ (0.0306) \end{array}$ & $\begin{array}{c} -0.1173\\ (0.0308) \end{array}$ \\
$\theta_{16}$ & & - & $\begin{array}{c} 0.2381\\ (0.1649) \end{array}$ & - & & - & - & - \\
$\theta_{17}$ & & - & $\begin{array}{c} 0.2825\\ (0.1619) \end{array}$ & - & & - & - & - \\
$\nu$ & & - & - & - & & $\begin{array}{c} 15.0003\\ (1.0914) \end{array}$ & $\begin{array}{c} 14.6560\\ (1.0627) \end{array}$ & $\begin{array}{c} 13.6868\\ (1.0247) \end{array}$ \\
\hline
\end{tabular}
\end{table}

As shown in Table~\ref{t:cacondefit}, the Wald test results confirm that the estimated parameter $\widehat{\beta}_1$ (also $\widehat{\beta}_2$ for KARMA with the cloglog link) is statistically significant at the 5\% level, supporting the correct identification of the seasonal component. The absence of the cosine term (in 5 out of 6 models) indicates that a phase adjustment was unnecessary to represent the annual seasonality. As a result, only the sine term was retained in most models, ensuring parsimony without compromising explanatory power.

\subsubsection*{Forecasting results and comparison}

\begin{figure}[!ht]
\centering
\subfigure[In-sample forecast]{\includegraphics[width=0.48\textwidth]{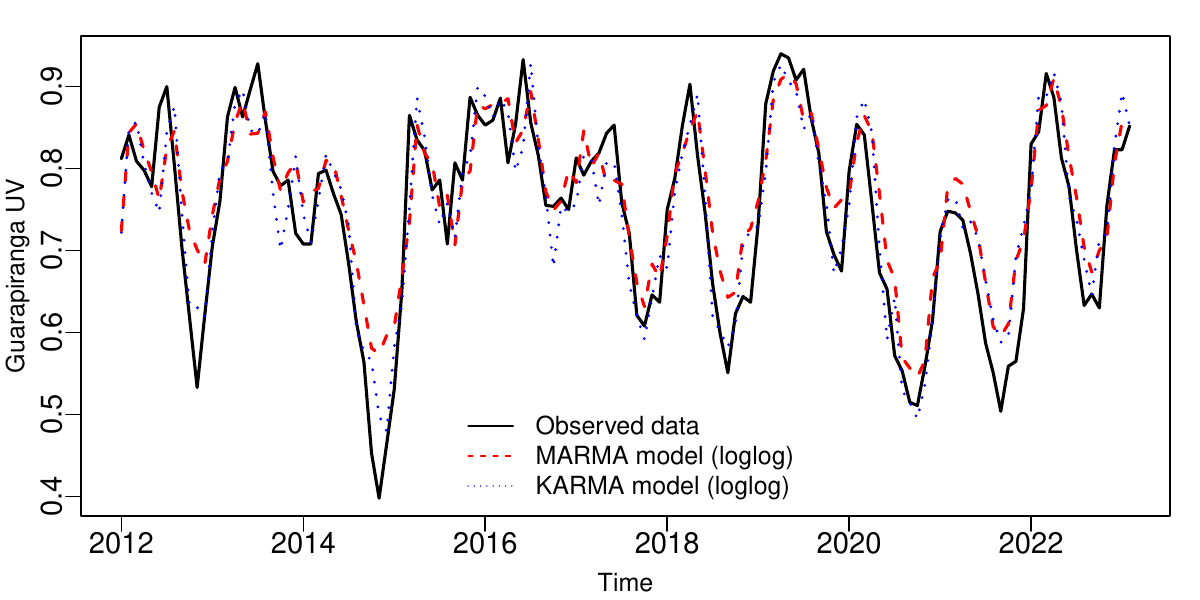}
\label{f:fit}}
\subfigure[Out-of-sample forecast]{\includegraphics[width=0.48\textwidth]{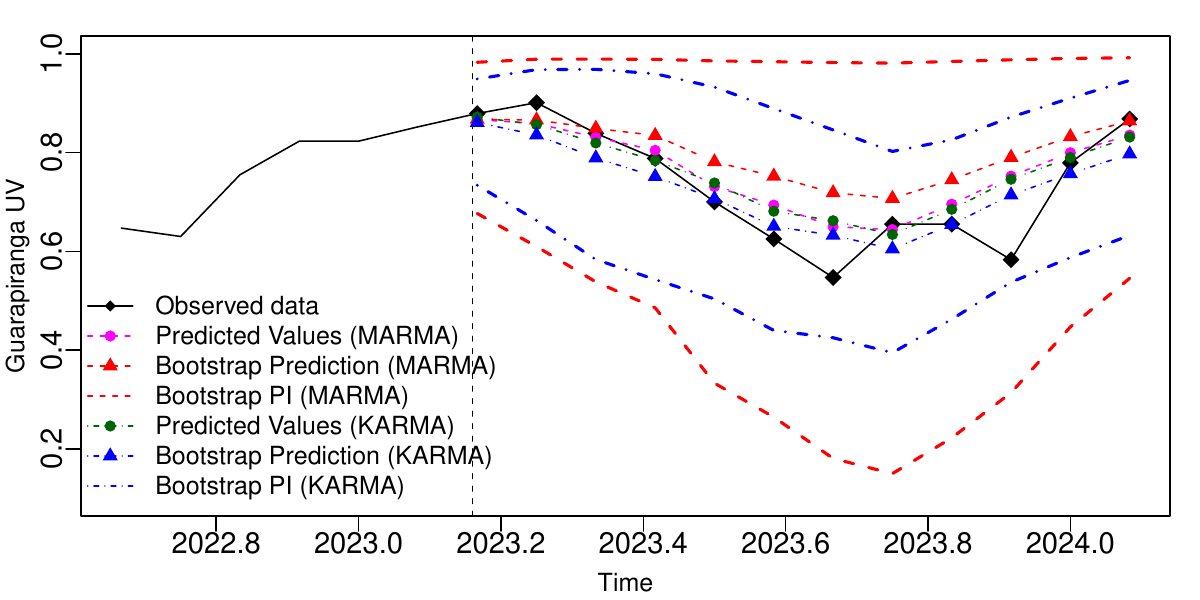}
 \label{f:prev}}
\caption{Observed time series, fitted values (in-sample-forecast) and out-of-sample forecasts obtained with the fitted MARMA model and the bootstrap method, for~$h=6$ steps ahead.}\label{f:adjusted}
\end{figure}

Figure~\ref{f:adjusted} illustrates the forecasting performance of the fitted MARMA and KARMA models using the loglog link function. Although the quantile residuals for the KARMA model did not pass the normality test, this model demonstrated the best out-of-sample forecasting performance among the KARMA models. Figure~\ref{f:fit} compares observed values within-sample predictions, highlighting the model's ability to capture key patterns in the data. Figure~\ref{f:prev} presents the out-of-sample forecasts for $h=12$ steps ahead, including both traditional and bootstrap predictions and the bootstrap prediction interval based on 10,000 bootstrap samples. Notably, both in-sample and out-of-sample predictions based on the model were very close, almost indistinguishable, indicating strong consistency in the model's performance. While the MARMA model's prediction intervals were wider than those of KARMA, indicating greater uncertainty, the strong agreement between the two forecasting methods and the well-contained confidence bounds further support the reliability of the predictions.

Table~\ref{t:forecast-guarapiranga-grouped} presents the out-of-sample forecast accuracy measures (RMSE, MAPE, and MDA) for the MARMA, KARMA, and Holt-Winters (HW) models across forecast horizons $h \in \{1, \cdots, 12\}$. For RMSE and MAPE, the best values for each measure and horizon are highlighted in \textbf{bold}. The RMSE and MAPE results reflect the models' predictive accuracy, where lower values indicate better performance. The MARMA model with the logit link function consistently outperformed other models for shorter horizons ($h \leq 3$), achieving the lowest RMSE and MAPE values. For longer horizons ($h \geq 4$), the KARMA model demonstrated superior performance across all link functions, particularly with the loglog link, which achieved the lowest RMSE and MAPE values for most horizons. In contrast, the Holt-Winters (HW) model exhibited significantly higher RMSE and MAPE values across all horizons, indicating weaker predictive accuracy than the MARMA and KARMA models.

\begin{table}[!ht]
\centering
\caption{Out-of-sample forecast accuracy measures RMSE, MAPE, and MDA for MARMA (M), KARMA (K), and Holt-Winters (HW) models with different link functions. For RMSE and MAPE, the best value for each measure is highlighted in \textbf{bold}.}\label{t:forecast-guarapiranga-grouped}\vskip 0.5\baselineskip
\renewcommand{\arraystretch}{1.3}
\renewcommand{\tabcolsep}{2pt}
\footnotesize
\begin{tabular}{l|ccccccccccccccc}
\cline{3-14}
\multicolumn{2}{c}{} & \multicolumn{12}{c}{Forecast Horizon ($h$)} \\
\hline
\multicolumn{2}{c}{Model} & 1 & 2 & 3 & 4 & 5 & 6 & 7 & 8 & 9 & 10 & 11 & 12\\
\hline
\multicolumn{14}{c}{RMSE}\\
\hline
\multirow{3}{*}{\rotatebox[origin=c]{90}{\footnotesize MARMA}} & logit & 0.0080 & \textbf{0.0210} & \textbf{0.0197} & 0.0266 & 0.0491 & 0.0685 & 0.0941 & 0.0914 & 0.0919 & 0.1104 & 0.1066 & 0.1021 \\
& loglog & 0.0122 & 0.0311 & 0.0258 & \textbf{0.0238} & \textbf{0.0255} & 0.0364 & \textbf{0.0515} & \textbf{0.0483} & \textbf{0.0475} & 0.0700 & 0.0670 & 0.0648 \\
& cloglog & 0.0133 & 0.0290 & 0.0238 & 0.0243 & 0.0424 & 0.0661 & 0.0974 & 0.0982 & 0.1022 & 0.1223 & 0.1186 & 0.1136 \\
\hline
\multicolumn{2}{c}{HW} & 0.0288 & 0.0281 & 0.0545 & 0.0879 & 0.1274 & 0.1676 & 0.2073 & 0.2111 & 0.2132 & 0.2195 & 0.2122 & 0.2038 \\
\hline
\multirow{3}{*}{\rotatebox[origin=c]{90}{\footnotesize KARMA}} & logit & 0.0143 & 0.0385 & 0.0340 & 0.0297 & 0.0315 & 0.0397 & 0.0600 & 0.0563 & 0.0561 & 0.0762 & 0.0728 & 0.0704 \\
& loglog & \textbf{0.0073} & 0.0319 & 0.0284 & 0.0247 & 0.0279 & \textbf{0.0343} & 0.0538 & 0.0509 & 0.0490 & \textbf{0.0694} & \textbf{0.0662} & \textbf{0.0643} \\
& cloglog & 0.0232 & 0.0430 & 0.0391 & 0.0352 & 0.0330 & 0.0361 & 0.0529 & 0.0495 & 0.0484 & 0.0705 & 0.0678 & 0.0651 \\
\hline
\multicolumn{14}{c}{MAPE}\\
\hline
\multirow{3}{*}{\rotatebox[origin=c]{90}{\footnotesize MARMA}} & logit & 0.0091 & \textbf{0.0204} & \textbf{0.0203} & 0.0282 & 0.0500 & 0.0755 & 0.1128 & 0.1119 & 0.1157 & 0.1409 & 0.1347 & 0.1241 \\
& loglog & 0.0139 & 0.0304 & 0.0234 & 0.0228 & \textbf{0.0272} & 0.0409 & \textbf{0.0620} & \textbf{0.0563} & \textbf{0.0569} & 0.0803 & 0.0754 & 0.0722 \\
& cloglog & 0.0151 & 0.0291 & 0.0211 & 0.0240 & 0.0424 & 0.0704 & 0.1127 & 0.1183 & 0.1272 & 0.1549 & 0.1492 & 0.1372 \\
\hline
\multicolumn{2}{c}{HW} & 0.0328 & 0.0316 & 0.0551 & 0.0884 & 0.1347 & 0.1911 & 0.2588 & 0.2715 & 0.2801 & 0.2984 & 0.2850 & 0.2664 \\
\hline
\multirow{3}{*}{\rotatebox[origin=c]{90}{\footnotesize KARMA}} & logit & 0.0163 & 0.0373 & 0.0339 & 0.0277 & 0.0330 & 0.0454 & 0.0717 & 0.0648 & 0.0669 & 0.0898 & 0.0835 & 0.0800 \\
& loglog & \textbf{0.0083} & 0.0288 & 0.0270 & \textbf{0.0216} & 0.0282 & \textbf{0.0385} & 0.0630 & 0.0591 & 0.0576 & \textbf{0.0798} & \textbf{0.0738} & \textbf{0.0712} \\
& cloglog & 0.0264 & 0.0444 & 0.0414 & 0.0372 & 0.0360 & 0.0430 & 0.0652 & 0.0577 & 0.0578 & 0.0811 & 0.0769 & 0.0722 \\
\hline
\multicolumn{14}{c}{MDA}\\
\hline
\multirow{3}{*}{\rotatebox[origin=c]{90}{\footnotesize MARMA}} & logit & 0.0000 & 0.0000 & 0.5000 & 0.6667 & 0.5000 & 0.4000 & 0.3333 & 0.4286 & 0.3750 & 0.3333 & 0.4000 & 0.4545 \\
& loglog & 0.0000 & 0.0000 & 0.5000 & 0.6667 & 0.7500 & 0.8000 & 0.6667 & 0.7143 & 0.6250 & 0.5556 & 0.6000 & 0.6364 \\
& cloglog & 0.0000 & 0.0000 & 0.5000 & 0.6667 & 0.7500 & 0.6000 & 0.5000 & 0.5714 & 0.5000 & 0.4444 & 0.5000 & 0.5455 \\
\hline
\multicolumn{2}{c}{HW} & 0.0000 & 1.0000 & 0.5000 & 0.3333 & 0.2500 & 0.2000 & 0.1667 & 0.2857 & 0.2500 & 0.2222 & 0.3000 & 0.3636 \\
\hline
\multirow{3}{*}{\rotatebox[origin=c]{90}{\footnotesize KARMA}} & logit & 0.0000 & 0.0000 & 0.5000 & 0.6667 & 0.7500 & 0.8000 & 0.6667 & 0.7143 & 0.6250 & 0.5556 & 0.6000 & 0.6364 \\
& loglog & 0.0000 & 0.0000 & 0.5000 & 0.6667 & 0.7500 & 0.8000 & 0.6667 & 0.7143 & 0.6250 & 0.5556 & 0.6000 & 0.6364 \\
& cloglog & 0.0000 & 0.0000 & 0.5000 & 0.6667 & 0.7500 & 0.8000 & 0.6667 & 0.7143 & 0.6250 & 0.5556 & 0.6000 & 0.6364 \\
\hline
\end{tabular}
\end{table}

The MDA results measure the models' ability to correctly predict the change direction in the forecasted values. Both MARMA and KARMA models generally achieved higher MDA values than the HW model, particularly for longer horizons ($h \geq 6$). The loglog link function consistently delivered the best MDA performance for both MARMA and KARMA models, highlighting its robustness in capturing directional changes. The HW model, however, struggled to maintain competitive MDA values, especially for longer horizons, further emphasizing its limitations in out-of-sample forecasting.

The comparison between in-sample and out-of-sample performance reveals important insights into model generalization. While the KARMA models fit the training data better, their out-of-sample performance varies across horizons, with MARMA models being more reliable for short-term forecasts. The cloglog link function achieved the best in-sample fit for MARMA but did not consistently dominate out-of-sample, highlighting a trade-off between in-sample fit and out-of-sample predictive accuracy. These results emphasize the importance of evaluating models both in-sample and out-of-sample to ensure robust and generalizable forecasting performance.  These findings align with the in-sample results, reinforcing the importance of selecting a model that balances accuracy, complexity, and computational cost. They also suggest that MARMA is a robust and reliable forecasting method for hydrological modeling.
\section{Conclusion}\label{sec:Conclusion}
The present work proposed the Matsuoka autoregressive moving average (MARMA) model for time series taking values in the interval $(0,1)$. The idea is to consider a GARMA specification based on the fact that the inference for the proposed model was conducted using partial maximum likelihood. The Matsoka distribution belongs to the exponential family in canonical form, so the asymptotic theory can be studied in detail. Inferential methods such as confidence intervals, residual analysis, model selection, and forecasting were also explored. In particular, a novel bootstrap-based method for constructing prediction intervals was introduced and studied. The proposed method can be applied to any GARMA model.

We provided a simulation study to assess the finite sample performance of the PMLE and the proposed method for constructing prediction intervals. We also assess goodness of fit tests, including a martingale difference test and normality tests for residuals. Overall, the results were satisfactory, indicating that the estimators behaved as expected. Furthermore, the overall finite-sample performance of the proposed bootstrap prediction confidence interval was good, as the empirical coverages were close to the nominal levels for all forecasting horizons.

The MARMA model was fitted to a Brazilian monthly useful water volume data in the empirical application, and a forecast exercise was employed. The MARMA estimates were compared with those obtained from KARMA and Holt-Winters models. Remarkably, HW was outperformed by both MARMA and KARMA models according to the usual out-of-sample accuracy measures considered in this work. With respect to KARMA, the out-of-sample performance of the MARMA forecasts varied across horizons. The MARMA models were more reliable for short-term forecasts, while the KARMA models performed better for long-term horizons. Turning to the in-sample performance, the KARMA model showed slightly better accuracy measures than MARMA. Overall, MARMA models exhibited a competitive performance with fewer parameters and significantly lower computational costs. These findings reinforce the practitioner to balance accuracy, complexity, and computational cost.

\FloatBarrier
\bibliographystyle{apalike}
\bibliography{marma}

\begin{thebibliography}{}

\bibitem[Bayer et~al., 2017]{Bayers}
Bayer, F.~M., Bayer, D.~M., and Pumi, G. (2017).
\newblock Kumaraswamy autoregressive moving average models for double bounded
  environmental data.
\newblock {\em Journal of Hydrology}, 555:385--396.

\bibitem[Bayer et~al., 2018]{Bayerseas}
Bayer, F.~M., Cintra, R.~J., and Cribari-Neto, F. (2018).
\newblock Beta seasonal autoregressive moving average models.
\newblock {\em Journal of Statistical Computation and Simulation},
  88(15):2961--2981.

\bibitem[Benaduce and Pumi, 2023]{helen}
Benaduce, H.~S. and Pumi, G. (2023).
\newblock {SYMARFIMA}: a dynamical model for conditionally symmetric time
  series with long range dependence mean structure.
\newblock {\em Journal of Statistical Planning and Inference}, 225:71--88.

\bibitem[Benjamin et~al., 2003]{Benjamin2003}
Benjamin, M., Rigby, R., and Stasinopoulos, D. (2003).
\newblock Generalized autoregressive moving average models.
\newblock {\em Journal of the American Statistical Association},
  98(461):214--223.

\bibitem[Casarin et~al., 2012]{casarin}
Casarin, R., Dalla~Valle, L., and Leisen, F. (2012).
\newblock Bayesian model selection for beta autoregressive processes.
\newblock {\em Bayesian Analysis}, 7(2):385--410.

\bibitem[Charles et~al., 2011]{charles}
Charles, A., Darn\'e, O., and Kim, J.~H. (2011).
\newblock Small sample properties of alternative tests for martingale
  difference hypothesis.
\newblock {\em Economics Letters}, 110(2):151--154.

\bibitem[Choi, 1999]{Choi}
Choi, I. (1999).
\newblock Testing the random walk hypothesis for real exchange rates.
\newblock {\em Journal of Applied Econometrics}, 14(3):293--308.

\bibitem[Consul and Jain, 1971]{consul}
Consul, P.~C. and Jain, G.~C. (1971).
\newblock On the log-gamma distribution and its properties.
\newblock {\em Statistische Hefte}, 12:100--106.

\bibitem[Cox et~al., 1981]{cox1981}
Cox, D.~R., Gudmundsson, G., Lindgren, G., Bondesson, L., Harsaae, E., Laake,
  P., Juselius, K., and Lauritzen, S.~L. (1981).
\newblock Statistical analysis of time series: Some recent developments [with
  discussion and reply].
\newblock {\em Scandinavian Journal of Statistics}, pages 93--115.

\bibitem[Dang et~al., 2020]{dang2020}
Dang, T.~D., Chowdhury, A.~K., and Galelli, S. (2020).
\newblock On the representation of water reservoir storage and operations in
  large-scale hydrological models: implications on model parameterization and
  climate change impact assessments.
\newblock {\em Hydrology and Earth System Sciences}, 24(1):397--416.

\bibitem[Fahrmeir, 1987]{Fahrmeir1987}
Fahrmeir, L. (1987).
\newblock Asymptotic testing theory for generalized linear models.
\newblock {\em Statistics}, 18(1):65--76.

\bibitem[Fokianos and Kedem, 1998]{Fokianos1998}
Fokianos, K. and Kedem, B. (1998).
\newblock Prediction and classification of non-stationary categorical time
  series.
\newblock {\em Journal of Multivariate Analysis}, 67:277--296.

\bibitem[Fokianos and Kedem, 2004]{Fokianos2004}
Fokianos, K. and Kedem, B. (2004).
\newblock Partial likelihood inference for time series following generalized
  linear models.
\newblock {\em Journal of Time Series Analysis}, 25(2):173--197.

\bibitem[Gradshteyn and Ryzhik, 2007]{grad}
Gradshteyn, I.~S. and Ryzhik, I.~M. (2007).
\newblock {\em Table of integrals, series, and products}.
\newblock Academic Press, 7 edition.

\bibitem[Grande et~al., 2022]{grande}
Grande, A.~F., Pumi, G., and Cybis, G.~B. (2022).
\newblock Granger causality and time series regression for modeling the
  migratory dynamics of {I}nfluenza into {B}razil.
\newblock {\em {SORT}}, 46(2).

\bibitem[Grassia, 1977]{Grassia}
Grassia, A. (1977).
\newblock On a family of distributions with argument between 0 and 1 obtained
  by transformation of the gamma and derived compound distributions.
\newblock {\em Australian Journal of Statistics}, 19(2):108--114.

\bibitem[Griffiths and Schafer, 1981]{griff}
Griffiths, D. and Schafer, C. (1981).
\newblock Closeness of {G}rassia's transformed gammas and the beta
  distribution.
\newblock {\em Australian Journal of Statistics}, 23(2):240--246.

\bibitem[Gross and Ligges, 2015]{nortest}
Gross, J. and Ligges, U. (2015).
\newblock {\em nortest: Tests for Normality}.
\newblock R package version 1.0-4.

\bibitem[Halliwell, 2021]{hell}
Halliwell, L.~J. (2021).
\newblock The {Log-Gamma} distribution and non-normal error.
\newblock {\em Variance}, 13.

\bibitem[Hogg and Klugman, 1984]{hogg}
Hogg, R.~V. and Klugman, S.~A. (1984).
\newblock {\em Loss distributions}.
\newblock John Wiley \& Sons.

\bibitem[Holt, 2004]{holt2004forecasting}
Holt, C.~C. (2004).
\newblock Forecasting seasonals and trends by exponentially weighted moving
  averages.
\newblock {\em International Journal of Forecasting}, 20(1):5--10.

\bibitem[Kalliovirta, 2012]{kall}
Kalliovirta, L. (2012).
\newblock Misspecification tests based on quantile residuals.
\newblock {\em The Econometrics Journal}, 15(2):358--393.

\bibitem[Kim, 2014]{vrtest}
Kim, J.~H. (2014).
\newblock {\em vrtest: Variance Ratio tests and other tests for Martingale
  Difference Hypothesis}.
\newblock R package version 0.97.

\bibitem[Maior and Cysneiros, 2018]{Maior}
Maior, V. and Cysneiros, F. (2018).
\newblock {SYMARMA}: a new dynamic model for temporal data on conditional
  symmetric distribution.
\newblock {\em Statistical Papers}, 59.

\bibitem[Matsuoka et~al., 2024]{mat}
Matsuoka, D.~H., Pumi, G., Torrent, H., and Valk, M. (2024).
\newblock A three-step approach to production frontier estimation and the
  {M}atsuoka's distribution.
\newblock {\em arXiv}, 2311.06086.

\bibitem[Palm et~al., 2021]{bruna}
Palm, B.~G., Bayer, F.~M., and Cintra, R.~J. (2021).
\newblock Prediction intervals in the beta autoregressive moving average model.
\newblock {\em Communications in Statistics - Simulation and Computation},
  0(0):1--22.

\bibitem[Prass and Pumi, 2022]{btsr}
Prass, T.~S. and Pumi, G. (2022).
\newblock {\em {BTSR:} Bounded Time Series Regression}.
\newblock R package version 0.1.0.

\bibitem[Prass et~al., 2024]{Prass}
Prass, T.~S., Pumi, G., Taufemback, C.~G., and Carlos, J.~H. (2024).
\newblock Positive time series regression models: theoretical and computational
  aspects.
\newblock {\em Computational Statistics}.
\newblock forthcoming.

\bibitem[Pumi et~al., 2021]{BARC}
Pumi, G., Prass, T.~S., and Souza, R.~R. (2021).
\newblock A dynamic model for double-bounded time series with chaotic-driven
  conditional averages.
\newblock {\em Scandinavian Journal of Statistics}, 48(1):68--86.

\bibitem[Pumi et~al., 2024]{uw}
Pumi, G., Prass, T.~S., and Taufemback, C.~G. (2024).
\newblock {Unit-Weibull} autoregressive moving average models.
\newblock {\em {TEST}}, 33:204--209.

\bibitem[Pumi et~al., 2020]{Pumi2020}
Pumi, G., Rauber, C., and Bayer, F.~M. (2020).
\newblock Kumaraswamy regression model with {A}randa-{O}rdaz link function.
\newblock {\em {TEST}}, 29:1051--1071.

\bibitem[Pumi et~al., 2019]{Pumi2017}
Pumi, G., Valk, M., Bisognin, C., Bayer, F.~M., and Prass, T.~S. (2019).
\newblock Beta autoregressive fractionally integrated moving average models.
\newblock {\em Journal of Statistical Planning and Inference}, 200:196--212.

\bibitem[{R Core Team}, 2023]{R}
{R Core Team} (2023).
\newblock {\em R: A Language and Environment for Statistical Computing}.
\newblock R Foundation for Statistical Computing, Vienna, Austria.

\bibitem[Reis et~al., 2024]{ufpe}
Reis, L. D.~R., Cordeiro, G.~M., and do~Carmo S.~Lima, M. (2024).
\newblock The unit gamma-{G} class: properties, simulations, regression and
  applications.
\newblock {\em Communications in Statistics - Simulation and Computation},
  53(8):3802--3829.

\bibitem[Rocha and Cribari-Neto, 2009]{Rocha2009}
Rocha, A.~V. and Cribari-Neto, F. (2009).
\newblock Beta autoregressive moving average models.
\newblock {\em {TEST}}, 18(3):529--545.

\bibitem[Thode, 2002]{thode}
Thode, H.~C. (2002).
\newblock {\em Testing for normality}.
\newblock CRC Press.

\bibitem[Winters, 1960]{winters1960forecasting}
Winters, P.~R. (1960).
\newblock Forecasting sales by exponentially weighted moving averages.
\newblock {\em Management Science}, 6(3):324--342.

\end{thebibliography}

\includepdf[pages=1-9]{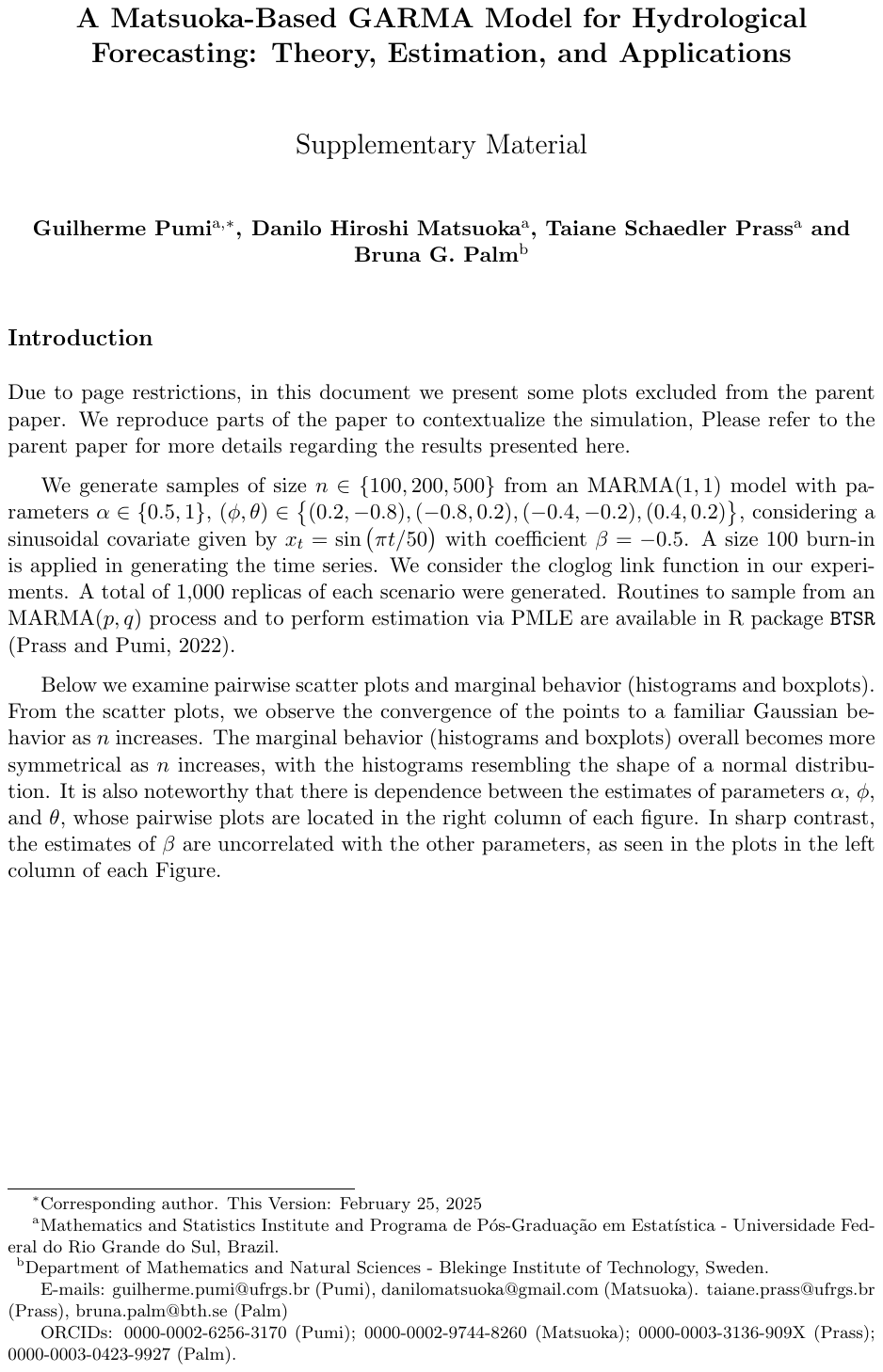}

\end{document}